\documentclass[sigconf,screen]{acmart}

\AtBeginDocument{%
  }

\usepackage{subfigure}
\usepackage{textcomp}
\usepackage{multirow}
\usepackage{pifont}
\usepackage{marvosym}
\definecolor{myellow1}{rgb}{0.94, 0.8, 0.0}
\colorlet{mygray}{white!15}
\colorlet{myellow}{lightgray!15}

\definecolor{g}{RGB}{189, 230, 205}
\definecolor{l}{RGB}{228,238,188}      
\definecolor{ll}{RGB}{255,248,197}

\acmSubmissionID{}
\renewcommand\footnotetextcopyrightpermission[1]{}
\begin{document}

\title{Mobile U-ViT: Revisiting large kernel and U-shaped ViT for efficient medical image segmentation}


\author{
    Fenghe Tang\textsuperscript{1,2}$^\dagger$ \quad
    Bingkun Nian\textsuperscript{3,4,5}$^\dagger$ \quad
    Jianrui Ding\textsuperscript{6}
    \quad
    Wenxin Ma\textsuperscript{1,2}
    \\
    Quan Quan\textsuperscript{7}
    \quad
    Chengqi Dong\textsuperscript{1,2}
    \quad
    Jie Yang\textsuperscript{3,4,5}
    \quad
    Wei Liu \textsuperscript{3,4,5} $^{\href{mailto:weiliucv@sjtu.edu.cn}{\textrm{\Letter}}}$
    \quad
    S. Kevin Zhou\textsuperscript{1,2} $^{\href{mailto:skevinzhou@ustc.edu.cn}{\textrm{\Letter}}}$
}
\affiliation{
    \textsuperscript{1}School of Biomedical Engineering, Division of Life Sciences and Medicine, \\ University of Science and Technology of China, Hefei, Anhui, 230026, P.R. China \\
    \textsuperscript{2}Center for Medical Imaging, Robotics, Analytic Computing \& Learning (MIRACLE),\\ Suzhou Institute for Advanced Research, USTC, 215123, P.R. China \\
    \textsuperscript{3} School of Automation and Intelligent Sensing, Shanghai Jiao Tong University \\
    \textsuperscript{4} Institute of Image Processing and Pattern Recognition, Shanghai Jiao Tong University \\
    \textsuperscript{5} Institute of Medical Robotics, Shanghai Jiao Tong University \\
    \textsuperscript{6} School of Computer Science and Technology, Harbin Institute of Technology, Harbin, China \\
    \textsuperscript{7} State Grid Hunan ElectricPower Corporation Limited Research Institute \\
    \country{Code: \url{https://github.com/FengheTan9/Mobile-U-ViT}}
}

\begin{abstract}
In clinical practice, medical image analysis often requires efficient execution on resource-constrained mobile devices. However, existing mobile models—primarily optimized for natural images—tend to perform poorly on medical tasks due to the significant information density gap between natural and medical domains. Combining computational efficiency with medical imaging-specific architectural advantages remains a challenge when developing lightweight, universal, and high-performing networks. To address this, we propose a mobile model called \textbf{Mobile} \textbf{U-}shaped \textbf{Vi}sion \textbf{T}ransformer (Mobile U-ViT) tailored for medical image segmentation. Specifically, we employ the newly purposed ConvUtr as a hierarchical patch embedding, featuring a parameter-efficient large-kernel CNN with inverted bottleneck fusion. This design exhibits transformer-like representation learning capacity while being lighter and faster. To enable efficient local-global information exchange, we introduce a novel Large-kernel Local-Global-Local (LGL) block that effectively balances the low information density and high-level semantic discrepancy of medical images. Finally, we incorporate a shallow and lightweight transformer bottleneck for long-range modeling and employ a cascaded decoder with downsample skip connections for dense prediction. Despite its reduced computational demands, our medical-optimized architecture achieves state-of-the-art performance across eight public 2D and 3D datasets covering diverse imaging modalities, including zero-shot testing on four unseen datasets. These results establish it as an efficient yet powerful and generalization solution for mobile medical image analysis. Code is available at \url{https://github.com/FengheTan9/Mobile-U-ViT}. 
\end{abstract}

\begin{teaserfigure}
  \includegraphics[width=\textwidth]{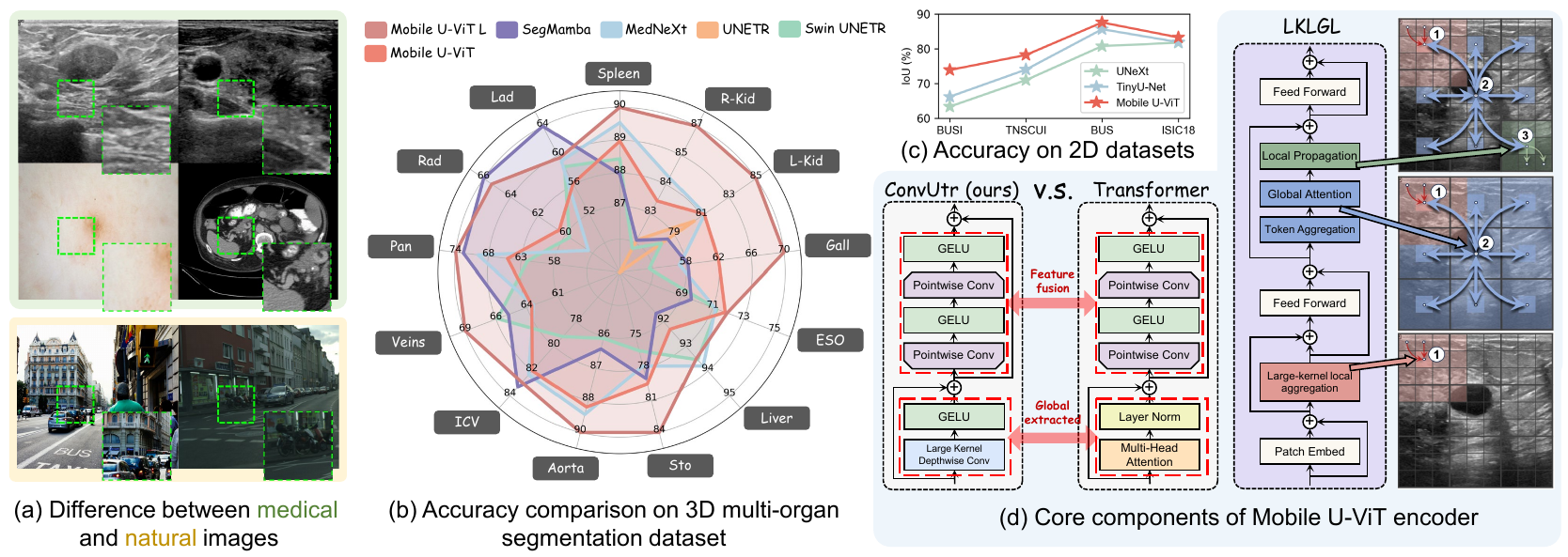}
       \captionof{figure}{The innovation and superior performance of Mobile U-ViT. (a) The information density gap between medical and natural images. Unlike natural images, medical images typically contain sparse local features, and their relevant information is often difficult to extract due to distributed noise and external artifacts. Since 2D lesion segmentation and 3D volume segmentation  rely on global context for inference and location, larger receptive fields are often needed to capture sufficient information. (b) The performance of different methods on 3D multi-organ segmentation datasets. Results demonstrate the superiority of Mobile U-ViT across different organs with diverse sizes. (c) The accuracy improvement of Mobile U-ViT on 2D datasets. The results validate the robustness of our method. (d) Mobile U-ViT encoder consists of two core components: ConvUtr and the Large-kernel Local-Global-Local (LKLGL) block. ConvUtr is CNN-based but shares Transformers' learning patterns, while LKLGL further refines both global and local understanding with high computational efficiency.
      }\label{fig:teaser}
\end{teaserfigure}

\maketitle

\renewcommand{\shortauthors}{Fenghe Tang et al.}

\section{Introduction}
Medical image segmentation is a critical and challenging task that can greatly enhance diagnostic efficiency by offering doctors objective and precise references for regions of interest~\cite{zhou2017deep}. However, this improved performance often comes with increased model size and inference latency~\cite{zhou2017deep,tang2024cmunext}.  In real-world clinical applications, such as point-of-care imaging intervention~\cite{poc} and real-time diagnosis~\cite{unext, egeunet, tang2024cmunext, fair}, timely execution on resource-constrained mobile devices is essential. To tackle the above efficiency challenge, mobile models with fewer parameters and lower FLOPs have attracted significant attention from researchers~\cite{mobilenet, mobilenetv2, mobilevit, emo}. In efficient model design, Convolutional Neural Networks (CNNs) offer a cost-effective approach for implementing lightweight backbones due to their high inference efficiency and strong inductive bias, which has led to notable progress in lightweight medical image segmentation~\cite{unext, egeunet, tinyunet, tang2024cmunext}. However, due to inherent limitations in capturing global context, pure CNN models struggle to achieve further breakthroughs in segmentation performance~\cite{limitations,hyspark,mambamim,llm4seg}.

Recently, mobile hybrid architectures trained on natural images leverage the inductive bias of CNNs~\cite{unet} and the global context learning capability of ViT~\cite{vits,hiend,ecamp,slide}, showing a great potential to break through the performance bottlenecks~\cite{mobilevit,edgevits,efficientvit}. However, their performance on medical visual tasks are remains limited. This is largely due to fundamental information density gap between medical and natural images: 
\textbf{(i) Sparse Local Information}.
Medical images generally contain less informative content within the same spatial window and exhibit higher similarity across adjacent regions (as illustrated in Fig.~\ref{fig:teaser} (a)). This sparse distribution of salient features makes it difficult for standard small-kernel convolutions to capture meaningful patterns, thus calling for larger receptive fields to aggregate sufficient contextual information.
\textbf{(ii) Blurred Boundaries and High Noise Levels}. Unlike natural images with clear object boundaries, medical images often present indistinct lesion edges and low contrast between target structures and surrounding tissues. Additionally, the presence of complex anatomical backgrounds and imaging artifacts further increases ambiguity (Fig.~\ref{fig:teaser}(a,d)). These characteristics demand models that can simultaneously emphasize subtle local details and suppress noise while leveraging global contextual cues to accurately differentiate between visually similar regions. We find that deep mobile networks designed for natural image tasks often fail to account for the two aforementioned challenges~\cite{mobilevit,edgevits,efficientvit}. This domain gap significantly hinder the performance of such models in medical image analysis.

In this paper, we aim to bridge the gap by designing a more effective and efficient mobile network specifically tailored for medical image segmentation:
\underline{For the issue (i)}, to strike a balance between enlarging the receptive field and  maintaining computational efficiency, we propose a Large kernel \textbf{Conv}olution block \textbf{U}sing \textbf{tr}ansformer-mode (ConvUtr) as the patch embeding, built upon depthwise separable convolutions (DSConv)~\cite{mobilenet}. The ConvUtr block composes large-kernel depth-wise convolution to extract global feature, followed by the inverted two point-wise convolutions to facilitate channel interaction. This design echoes the modeling pattern of Transformer\cite{transfomer} (as illustrated in Fig.~\ref{fig:teaser}(d)). Notably, compared to vanilla ViT, ConvUtr significantly reduces the number of parameters, making it a lighter and faster alternative for mobile applications. 
\underline{For the issue (ii)}, we integrate the \textbf{L}arge-\textbf{K}ernel \textbf{L}ocal-\textbf{G}lobal-\textbf{L}ocal (LKLGL) module. This component is designed to reduce semantic ambiguity and enhance the integration of local and global features by facilitating structured information flow: it first performs local feature aggregation (Red), then enables global context exchange (Blue) efficiently, and finally redistributes refined information locally (Green), as illustrated in Fig.~\ref{fig:teaser}(d). The inclusion of token aggregation operations within the LKLGL module reduces the number of tokens, enabling more efficient long-range computations.
Furthermore, we build a lightweight \textbf{cascaded decoder} with downsampled skip-connections, enables efficient and fast decoding while facilitating alignment and dense prediction of both fine-grained local details and high-level global semantics.

Through extensive experiments, our medical-optimized architecture has shown state-of-the-art (SOTA) performance across eight public 2D and 3D datasets from various imaging modalities. Its zero-shot generalization capability has also been validated, reinforcing its effectiveness. Because of the  significantly reduced resource demands and superior performance, Mobile U-ViT serves as both an efficient and powerful solution for mobile medical image analysis. 

To sum up, we propose a novel and effective hybrid light-weight network called  \textbf{Mobile U-ViT} to tackle challenges for mobile medical imaging. Mobile U-ViT achieves SOTA results with the help of three key parts: 
\begin{itemize}
    \item \textbf{ConvUtr} – a lightweight, Transformer-inspired CNN backbone that efficiently compresses medical images from the sparse pixel space into a compact latent representation;
    \item \textbf{Large-Kernel Local-Global-Local (LKLGL)} modules – designed to enable efficient interaction between local and global information flows for robust feature refinement; 
    \item \textbf{Cascaded decoder} with downsampled skip-connections, designed to align local and global features effectively, thereby facilitating accurate and efficient dense prediction.
\end{itemize}

\begin{figure*}[!t]
    \centering
    \includegraphics[width=0.99\linewidth]{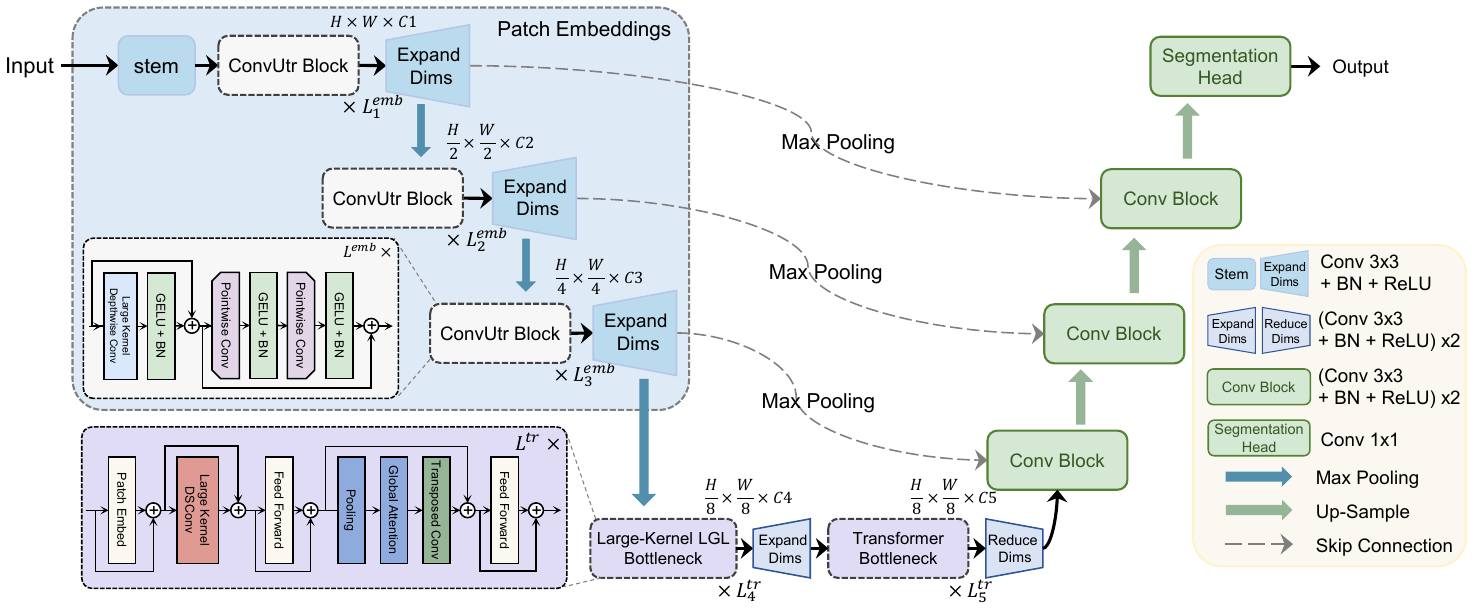}
    \caption{The architecture of Mobile U-ViT. The encoder is divided into 5 stages: first 3 stages are ConvUtr block with CNN structure, the 4-th stage is Large-kernel LGL blocks. Each decoder block combines an upsampling block and convolution block with downsampled skip-connection for feature fusion.}
    \label{fig:MobileUtr}
\end{figure*}

\section{Method}
\label{sec:Method}
The overall architecture of Mobile U-ViT is shown in Fig.~\ref{fig:MobileUtr}. Following previous works~\cite{mobilevit,transunet,cmunet}, we use hybrid segmentation model. The encoder (Sec.~\ref{sec:encoder})  integrates the CNN-based ConvUtr with Large-Kernel Local-Global-Local (LKLGL) modules to generate embeddings for the transformer layer,  while the decoder (Sec.~\ref{sec:decoder}) adopts a cascaded upsampling structure enhanced by downsampled skip-connections for efficient segmentation prediction.

\subsection{Efficient hybrid encoder}
\label{sec:encoder}

\subsubsection{ConvUtr.}

Depthwise separable convolution (DSConv) is widely used in lightweight CNNs, but its design — primarily optimized for natural images — poses limitations in medical imaging tasks. Medical images often exhibit significant variations in the position and shape of lesions and organs, which DSConv struggles to capture due to its inherently limited receptive field. Therefore, it is important to balance model efficiency with representational capacity by expanding the receptive field without increasing model size. We design a medically tailored lightweight component that enhances semantic representation while remaining computationally efficient. 

To achieve a lightweight patch embedding without compromising performance, our ConvUtr block comprises large kernel DepthwiseConv$(\cdot)$ with two inverted  PointwiseConv$(\cdot)$ to mimic the learning mechanism in ViT, enabling efficient long-range dependency modeling within a convolutional framework. Given an image $X \in \mathcal{R}^{H \times W \times 3}$, we attempt to utilize an encoder composed by ConvUtr blocks to output embeddings $X_e$ for LKLGL block (detaited in Sec.~\ref{sec:LKLGL}). The definition of ConvUtr block is as follows:
\begin{equation}
    \begin{aligned}
    Y_l &= \text{BN}\left(\sigma\left\{\text{DepthwiseConv}\left(X_l;K_{large}\right)\right\}\right)+X_l, \\
    Z_l &= \text{BN}\left(\sigma\left\{\text{PointwiseConv}\left(Y_l\right)\right\}\right), \\
    X_{l+1} &= \text{BN}\left(\sigma\left\{\text{PointwiseConv}\left(Z_l\right)\right\}\right) + Y_l, \\
    \end{aligned}
\end{equation}
where $X_l$ represents the output feature map of the $l$-th layer in the ConvUtr block, $K_{large}$ is the large kernel, $Y_l$ and $Z_l$ are the intermediate variables, $\sigma$ denotes the GELU function~\cite{gelu}, and $BN$ denotes batch normalization. The block is stacked various times to allow hierarchical modeling.
The detailed configuration, including stacked times ($L_1^{emb}$, $L_2^{emb}$, $L_3^{emb}$), kernel size (K1, K2, K3), and the number of channels (C1, C2, C3) of each block, are shown in Tab.~\ref{Table:variants}.

Notably, we use max-pooling for downsampling at the end of each block. This operation effectively reduces background noise and help to sharpen the poorly defined boundary, benefiting segmentation considering the characteristics of medical images.

\subsubsection{Large-kernel Local-Global-Local (LKLGL) block}
\label{sec:LKLGL}

To mitigate semantic ambiguity and strengthen the integration of local and global representations, we introduce Large-kernel LGL (LKLGL) blocks to optimize information before the transformer layer. Inspired by ~\cite{edgevits}, each LKLGL has four steps: 1) capturing rich local information through a large-kernel \textit{DSConv} block; 2) utilizing pooling operation to aggregate tokens for efficient long-range modeling, 3) employing attention mechanism~\cite{transfomer} for effective global information exchange, 4) applying transposed convolution to effectively distribute the flow at the local level. This process is formulated as:
\begin{equation}
    \begin{aligned}
        X &= \text{DSConv}(\text{Norm}(X_{in});K_{large}) + X_{e}, \\
        Y &= \text{FFN}(\text{Norm}(X)) + X, \\
        Z &= \text{TransConv}(\text{Pool}(\text{Attn}(\text{Norm}(Y)))) + Y,\\
        X &_{\text{out}} = \text{FFN}(\text{Norm}(Z)) + Z, \\
    \end{aligned}
    \label{eq:lgl}
\end{equation}
where $X_{e}$ is the output of ConvUtr blocks and $X_{out}$ is the input to the Transformer. Pool$(\cdot)$ is the average pooling with ratio $p$. TransConv$(\cdot)$ is the transposed convolution. 
We expand the kernel size $K_{large}$ of DSConv$(\cdot)$ with a larger receptive field, enabling rich feature aggregation.
During the whole process, the pooling operation reduces the token number by $p^2$, leading to improved computational efficiency.

\subsubsection{Computational Complexity}

\noindent\textbf{ConvUtr.} For a conventional convolutional layer, it takes an input $X_{in}$ of size $h \times w \times d_{i}$ and applies a convolutional kernel $K \in \mathcal{R}^{k \times k \times d_{i} \times d_{j}}$, producing an output of size  $h \times w \times d_{j}$. The computational complexity of conventional convolution layer is given by: $\mathcal{O}_{\text{Conv}} = h \times w \times d_{i} \times d_{j} \times k^{2}$. Our ConvUtr serves as an efficient and effective alternative to improve medical segmentation performance, but only cost: $\mathcal{O}_{\text{ConvUtr}} = h \times w \times d_{i} \times ( k^{2} + 2 \times d_{j})$, which is the sum of the depthwise and two pointwise convolutions.

\noindent\textbf{LKLGL.} Vanilla Transformers suffer from substantial computational overhead, with a complexity $\approx \mathcal{O}(N^2)$, where $N = h \times w$ represents the number of visual tokens. The LKLGL introduces a novel token aggregation mechanism with pooling ratio $p$ that effectively reduces the computational complexity $\approx \mathcal{O}(N^2/p^4)$, while preserving long-range modeling capacity.

\subsection{Cascade decoder with downsampled skip-connections}
\label{sec:decoder}

\subsubsection{Decoder Structure}

Following previous works~\cite{unet,cmunet,nnunet,swinunet,transunet,uctransnet}, we adopt a progressive cascade decoding approach with residual connections from the encoder. This approach consists of multiple stages, each comprising an upsampling layer, a convolution layer, a batch normalization layer, and a ReLU activation function. Detailed structure description can be found in the Appendix.

During decoding, aligning low-level CNN features with high-level Transformer output representations is challenging due to their semantic disparity. To address this, we introduce downsampled skip-connections that operate on encoder features and decoder features with different resolution. Beyond semantic alignment, this design filters out redundant information (\textit{e.g.} background tissue, external noises) in encoder features, exaggerating necessary boundary information for segmentation.

\section{Experiments}
\label{sec:Experiment}

\subsection{Experiment Setting}
\textbf{Dataset}: We select six public 2D datasets and two public 3D datasets for evaluation, spanning four medical image modalities: CT (Synapse\footnote{\url{https://www.synapse.org/\#!Synapse:syn3193805/files/}} and BTCV~\cite{btcv} for multi-oragn segmentation and KiTS'23~\cite{kits23} for kidney tumor segmentation), ultrasound (BUS~\cite{DATA_BUS} and BUSI~\cite{DATA_BUSI} for breast lesion segmentation, TNSCUI\footnote{\url{https://tn-scui2020.grand-challenge.org/Dataset/}} for thyroid lesion segmentation), dermoscopy (ISIC2018~\cite{DATA_ISIC2018}), and polypscopy (Kvasir~\cite{kvasir}) images. Additionally, to validate the generalization ability of our method, we conducted zero-shot experiments on the BUS~\cite{DATA_BUS}, TUCC\footnote{\url{https://aimi.stanford.edu/datasets/thyroid-ultrasound-cine-clip}}, PH2~\cite{ph2}, and CVC-300~\cite{cvc300}. Detailed dataset information and partition description can be found in Appendix.

{\noindent\textbf{{Evaluation metrics:}} 
Following~\cite{unext,tinyunet}, we use below metrics to measure computational efficiency: the number of parameters (Params) measures the model's size; Frames per second (FPS) indicates the inference speed; GFLOPs (giga floating-point operations) quantify the computational complexity. For segmentation evaluation, following~\cite{unext,cmunet,tinyunet}, we utilize Intersection over Union (IoU) and F1 score for BUS, BUSI, TNSCUI and ISIC, while adopting the Dice, Jaccard for Synapse and Dice, Hausdorff Distance (HD95), and Average Surface Distance (ASD) for BTCV, following~\cite{unetr,swinunetr,mednext,transunet}. 

{\noindent\textbf{{Comparison methods:}}}
We selected 22 recent SOTA 2D and 3D medical three types of segmentation models for comparison: (1) \textit{2D heavy-weight medical image networks}, including U-Net~\cite{unet}, CMU-Net~\cite{cmunet}, nnUNet~\cite{nnunet} TransUnet~\cite{transunet}, Swin-Unet~\cite{swinunet}, MissFormer~\cite{missformer}, and UCTransNet~\cite{uctransnet}; (2)\textit{ light-weight natural image networks}, including MobileViT~\cite{mobilevit}, EdgeViT~\cite{edgevits}, RepViT~\cite{repvit}, EMO~\cite{emo}, EfficientViT~\cite{efficientvit}, and UniRepLKNet~\cite{unireplknet}; (3) \textit{light-weight medical image models}, including MedT~\cite{medt}, UNeXt~\cite{unext}, EGE-Unet~\cite{egeunet}, ERDUnet~\cite{erdunet} and TinyU-Net~\cite{tinyunet}; 3D heavy-weight medical image networks: UNETR~\cite{unetr}, Swin UNETR~\cite{swinunet}, MedNeXt~\cite{mednext} and SegMamba~\cite{segmamba}.

\noindent\textbf{Implementation details:}
The configuration for ConvUtr is listed in Tab.~\ref{Table:variants}. In LKLGL blocks, $K_{large}$ for DSConv is 9, pooling ratio is 2, and kernel size for transposed convolution is 2. The loss function for optimization
$\mathcal{L}_{seg}$ between the predicted $\hat{y}$ result and ground truth $y$ is defined as a combination of Binary Cross Entropy $\mathcal{L}_{BCE}$ and Dice loss  $\mathcal{L}_{Dice}$: $    \mathcal{L}_{seg} = 0.5 \times \mathcal{L}_{BCE} + \mathcal{L}_{Dice}.$

The input images are resize to 256$\times$256. For fair comparison, we follow the same training parameter settings and data augmentation from prior 2D segmentation methods (BUS, BUSI, and TNSCUI followed as~\cite{cmunet,unext} and Synapse followed as~\cite{transunet}). For 3D datasets, the data preprocessing strategy is the same as UNETR~\cite{unetr}. All the experiments are conducted using a single NVIDIA GeForce RTX4090 GPU. More details can be found in Appendix.

\renewcommand{\multirowsetup}{\centering}  
\begin{table}[t!]
\caption{Mobile U-ViT variants (Base / Large). \label{Table:variants}}
    \begin{tabular}{l  c | l  c | l  c}
    \hline
    \multicolumn{2}{c|}{Channels} & \multicolumn{2}{c|}{Block lengths} & \multicolumn{2}{c}{Kernel size} \\
    \hline
    C1 & \multicolumn{1}{c|}{16 / 32} & $L_1^{emb}$ & 1 / 1 &  K1 & 3 / 3 \\
    C2 & \multicolumn{1}{c|}{16 / 32} & $L_2^{emb}$  & 1 / 1 & K2 & 3 / 3\\
    C3 & \multicolumn{1}{c|}{32 / 64} & $L_3^{emb}$  & 3 / 3 & K3 & 7 / 7\\
    C4 & \multicolumn{1}{c|}{64 / 128} & $L_4^{tr}$  & 3 / 3 \\
    C5 & \multicolumn{1}{c|}{128 / 256} & $L_5^{tr}$  & 3 / 4 \\
    \hline
    \end{tabular}
    \vspace{-4mm}
\end{table}

\renewcommand{\multirowsetup}{\centering}  
\begin{table*}[t!]

\caption{Results on 2D Medical Datasets ($\text{Mean}_{\text{Std}}\%$).  Best results are highlighted as \colorbox{g}{\bf \!first\!}, \colorbox{l}{\!second\!} and \colorbox{ll}{\!third\!}. \colorbox{myellow}{Gray} background indicate hybrid architectural method. $\dagger$ denotes deep supervision. \label{tab:ExOnUltrasound}}
\resizebox{1\linewidth}{!}
{
\begin{tabular}{l rrr|ccc|c|c|c}
\hline 

\multirow{3}{*}{Network} & \multirow{3}{*}{Params↓} & \multirow{3}{*}{FPS↑} & \multirow{3}{*}{GFLOPs↓} & \multicolumn{6}{c}{IoU (\%)} \\
\cline{5-10}
& & & & \multicolumn{3}{c|}{Ultrasound} &  \multicolumn{1}{c|}{Dermoscopy} &  \multicolumn{1}{c|}{Polypscopy} & \multirow{2}{*}{Avg} \\
\cline{5-9}
&&&& \multicolumn{1}{c}{BUS  } & \multicolumn{1}{c}{BUSI  } & \multicolumn{1}{c|}{TNSCUI  } & \multicolumn{1}{c|}{ISIC } & \multicolumn{1}{c|}{Kvasir } &  \\
\cline{5-10}
\hline
\multicolumn{3}{l}{\textit{{\color{Gray} heavy-weight medical network}}} &&&&&& \\

\rowcolor{mygray} U-Net (MICCAI'15)~\cite{unet} & 34.52\,M & 139.32 & 65.52 & $86.73_{1.41}$  & $68.61_{2.86}$ & $75.88_{0.18}$ & $82.18_{0.87}$ & $86.52$ & 79.98\\
\rowcolor{mygray} CMU-Net (ISBI'23)~\cite{cmunet} & 49.93\,M & 93.19  & 91.25 & $87.18_{0.59}$  & $71.42_{2.65}$  & $77.12_{0.49}$ & $82.16_{1.06}$ &  {89.02} & 81.38 \\
\rowcolor{mygray} nnUNet (NM'21)~\cite{nnunet}  & 26.10\,M   & ---    & 12.67   & {$87.51_{1.01}$}  & {$72.11_{3.51}$}  & \cellcolor{g}{\textbf{78.99}$_{0.14}$} & \cellcolor{ll}{$83.31_{0.59}$} & $85.60$ & {81.43} \\
Swin-Unet (ECCV'22)~\cite{swinunet} & 27.14\,M  & 392.21 & 5.91  & $85.27_{1.24}$ &  $63.59_{4.96}$  & $75.77_{1.29}$ & $82.15_{1.44}$ & $73.58$ & 76.07  \\ 
\rowcolor{myellow} TransUnet (ArXiv'21)~\cite{transunet} & 105.32\,M & 112.95 & 38.52 & {$87.35_{1.24}$}  & $71.39_{2.37}$ & $77.63_{0.14}$ & $83.17_{1.25}$ & $85.07$ & 80.92 \\  
\rowcolor{myellow} UCTransNet (AAAI'22)~\cite{uctransnet} & 66.24\,M & 57.11 & 32.98 & $85.98_{1.34}$ & $68.47_{2.61}$  & $73.87_{0.24}$ & $82.67_{0.40}$ & {$87.97$} & 79.79\\
 MissFormer (TMI'22)~\cite{missformer} & 35.45\,M & 151.46 & 7.25 & $81.92_{1.20}$ 
& $63.29_{2.99}$  & $68.26_{2.47}$ & $80.99_{0.28}$ & $86.37$ & 76.16 \\
EMCAD (CVPR'24)~\cite{rahman2024emcad} & 26.76\,M & 57.45 & 5.60 & \cellcolor{ll}{$87.81_{1.21}$} & $71.09_{2.81}$ & $77.48_{0.45}$ & $82.91_{0.38}$ & \cellcolor{l}{88.45} & 81.55\\
\hline                                    
\multicolumn{3}{l}{\textit{{\color{Gray} light-weight natural network}}} &&&&&& \\

\rowcolor{myellow} MobileViT-s (ICLR'22)~\cite{mobilevit} & 10.66\,M  & 301.11 & 2.13  & $82.57_{1.38}$ & $64.28_{3.78}$ & $71.64_{0.28}$ & $80.12_{0.42}$ & $84.21$ & 76.56 \\
\rowcolor{myellow} EdgeViT-s (ECCV'22)~\cite{edgevits} & 16.49\,M  & 291.44 & 2.71  & $81.32_{1.23}$ & $61.12_{3.69}$ & $68.74_{0.29}$ & $79.06_{0.56}$ & $81.58$ & 74.36 \\

\rowcolor{myellow} EMO-6m (ICCV'23)~\cite{edgevits} & 9.14\,M  & 329.75 & 1.85  & $84.89_{0.86}$  & $67.50_{4.26}$  & $74.02_{0.39}$ & $81.31_{0.73}$ & $83.91$ & 78.32  \\ 
\rowcolor{myellow} EfficientViT-b1 (ICCV'23)~\cite{efficientvit}  & 4.79\,M & 181.79 & 9.18 & $84.81_{0.97}$  & $67.21_{2.72}$ & $71.18_{4.92}$ & $82.22_{0.56}$ & 83.43 & 77.77 \\
\rowcolor{myellow} RepViT-m3 (CVPR'24)~\cite{repvit}   & 14.37\,M  & 238.90 & 2.79   & $76.51_{1.38}$ & $56.07_{3.51}$ & $66.21_{0.66}$ & $78.34_{0.61}$ & $81.55$ & 71.73 \\
UniRepLKNet (CVPR'24)~\cite{unireplknet}  &5.83\,M & 178.36 & 9.39 & $84.13_{1.32}$  
&$ 65.26_{1.72}$  & $67.73_{0.92}$ & $81.64_{0.24}$ & $85.30$ & 76.81 \\
\hline 
\multicolumn{3}{l}{\textit{{\color{Gray} light-weight medical network}}} &&&&&& \\

\rowcolor{myellow} MedT (MICCAI'21)~\cite{medt}  & 1.37\,M   & 22.97  & 2.40  & $80.81_{2.77}$  & $63.36_{1.56}$  & $71.00_{2.68}$ & $81.79_{0.94}$  & $79.70$ & 75.33 \\
\rowcolor{mygray} UNeXt (MICCAI'22)~\cite{unext}   & 1.47\,M   & 650.48 & 0.58  & $84.73_{1.23}$  & $65.04_{2.71}$ & $71.04_{0.17}$ & $82.10_{0.88}$ & $84.78$ & 77.53 \\
\rowcolor{mygray} EGE-Unet (MICCAI'23)~\cite{egeunet}  & 0.072\,M  & $303.08$ & $0.045$ & $84.72_{1.28}$  & $58.90_{2.97}$ & $74.47_{0.43}$ & $82.19_{1.31}$ & $77.41$ & 75.53 \\
\rowcolor{mygray} ERDUnet (TCSVT'23)~\cite{erdunet} & 10.21\,M & 137.60 & 10.29 & $85.92_{1.50}$  & $68.67_{1.90}$ & $76.02_{0.40}$ & $82.03_{0.20}$ & $84.93$ & 75.53 \\
CMUNeXt (ISBI'24)~\cite{tang2024cmunext} & 3.14\,M & 471.21 & 7.41  & $87.27_{0.95}$ & $71.20_{2.24}$ & $76.96_{0.21}$ & $82.47_{0.47}$ & 87.17 & 81.01 \\
\rowcolor{mygray} TinyU-Net (MICCAI'24)~\cite{tinyunet} & 0.48\,M & 359.50 & 1.66 & $85.67_{0.99}$ & $66.21_{2.70}$ & $74.03_{0.20}$ & $81.95_{0.30}$ & $85.47$ & 78.66  \\

\hline
\rowcolor{myellow} Mobile U-ViT (Ours)  & 1.39\,M   & 326.24 & 2.51  & $87.28_{0.83}$  & {$72.88_{2.72}$} & {$77.70_{0.50}$} & {$83.23_{0.39}$} & 87.28 & {81.67}\\
\rowcolor{myellow} Mobile U-ViT$^\dagger$ (Ours)  & 1.39\,M & 292.32 & 2.52 & \cellcolor{l}{$87.86_{0.80}$} & \cellcolor{ll}{$73.34_{2.69}$} & $77.90_{0.41}$ & $83.22_{0.40}$ & {87.97} & \cellcolor{ll}{82.06}\\
\rowcolor{myellow} Mobile U-ViT-L (Ours)  & 7.88\,M  & 279.42  & 3.70 & 87.63$_{0.91}$ &  \cellcolor{g}{\textbf{73.91}$_{2.65}$} & \cellcolor{ll}{$78.24_{0.38}$} & \cellcolor{l}{83.31}$_{0.36}$ & \cellcolor{g}{\textbf{89.07}} & \cellcolor{g}{\textbf{82.43}} \\
\rowcolor{myellow} Mobile U-ViT-L$^\dagger$ (Ours)  & 7.89\,M & 228.59 & 3.71 & \cellcolor{g}{\textbf{87.87}$_{0.75}$} & \cellcolor{l}{$73.80_{2.34}$} & \cellcolor{l}{$78.62_{0.34}$} & \cellcolor{g}{\textbf{83.53}$_{0.43}$} & \cellcolor{ll}{88.10} & \cellcolor{l}{82.38}\\
\hline
\end{tabular}
}

\end{table*}

\begin{figure*}[t!]
  \centering
   \includegraphics[width=0.98\linewidth]{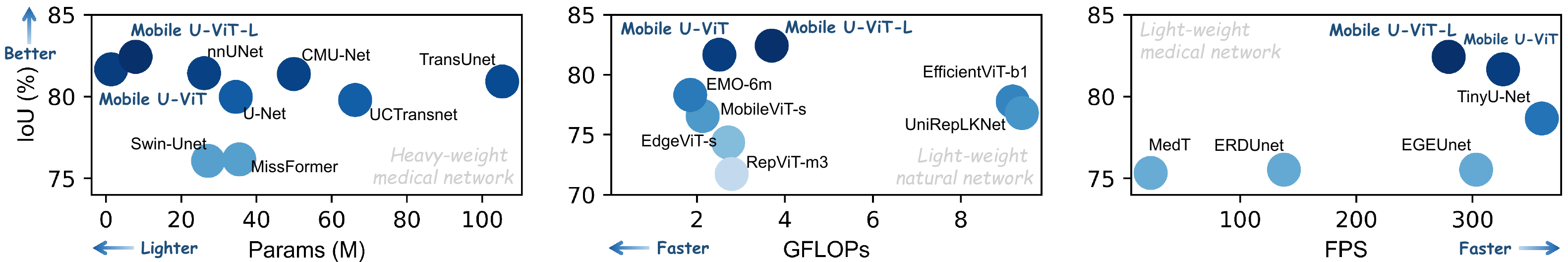}
   \caption{Performance \textit{v.s.} Params, GFLOPs and FPS across heavy- and light-weight medical and natural networks. We report the average IoU of five 2D datasets across three modalities.}
      \label{fig:performance}
\end{figure*}

\renewcommand{\multirowsetup}{\centering}  
\begin{table*}[t!]
\caption{Results on slice-level CT Dataset (Synapes). Best results are highlighted as \colorbox{g}{\bf \!first\!}, \colorbox{l}{\!second\!} and \colorbox{ll}{\!third\!}. \colorbox{myellow}{Gray} backgrounds indicates hybrid architectural method. \label{tab:ct}}
\resizebox{1\linewidth}{!}
{
\begin{tabular}{l  rrr|cc|cccccccc   }
\hline 

\multirow{2}{*}{Network} &  \multirow{2}{*}{Params↓} & \multirow{2}{*}{FPS↑} & \multirow{2}{*}{GFLOPs↓} & \multicolumn{2}{c|}{Avg. Performance}
& \multicolumn{8}{c}{Performance for each organ (Jaccard\%↑)} \\
\cline{5-7}\cline{7-14}
&&&&Jaccard\%↑ &Dice\%↑& Aorta & Gall. & Kid (L) & Kid (R) & Liver & Panc. & Spl. & Stom. \\
\hline
\multicolumn{3}{l}{\textit{{\color{Gray} heavy-weight medical network}}}&&&&& \\

Swin-Unet~\cite{swinunet} & 27.14\,M  & 392.21 & 5.91 & 62.42 & 74.13 
          & 70.43 & 35.73 & 68.58 & 61.64 & 88.52 & 37.01 & 79.48 & 57.97\\
\rowcolor{myellow} TransUnet~\cite{transunet} & 105.32\,M & 112.95 & 38.52 & \cellcolor{l}{68.33} & \cellcolor{ll}{79.12} 
          & 78.47 & 41.83 & \cellcolor{g}{\textbf{75.26}} & \cellcolor{g}{\textbf{72.59}} & \cellcolor{l}{89.90} & \cellcolor{l}{44.87} & \cellcolor{ll}{80.07} & \cellcolor{l}{63.68} \\
\rowcolor{myellow} UCTransNet~\cite{uctransnet} & 66.24\,M & 57.11 & 32.98 & 66.57 & 77.55  
          & 78.97 & \cellcolor{l}{46.37} & 71.14 & 64.16 & \cellcolor{ll}{88.88} & 41.65 & \cellcolor{l}{81.01} & 60.37\\
MissFormer~\cite{missformer} & 35.45\,M & 151.46 & 7.25 & 60.24 & 72.53 
          & 70.40 & 43.07 & 66.26 & 54.94 & 86.07 & 29.37 & 79.47 & 52.34\\
\hline
\multicolumn{3}{l}{\textit{{\color{Gray} light-weight natural network}}}&&&&& \\
\rowcolor{myellow} MobileViT-s~\cite{mobilevit} & 10.66\,M & 291.44 & 2.13 & 41.66 & 55.57  
          & 24.63 & 16.58 & 50.94 & 45.85 & 77.74 & 17.90 & 58.80 & 40.83\\
\rowcolor{myellow} EdgeViT-s~\cite{edgevits} & 10.49\,M  & 291.44 & 2.71 & 36.86 & 50.68  
          & 21.97 & 12.23  & 41.42 & 38.96 & 76.43 & 16.00 & 50.71 & 37.16\\
\rowcolor{myellow} RepViT-m3~\cite{repvit}  & 14.37\,M  & 238.90 & 2.79 & 34.07 & 47.61 
          & 20.71 & \ \;7.16 & 33.56 & 37.28 & 74.33 & 13.28 & 48.13 & 38.09\\
\rowcolor{myellow} EMO-6m~\cite{emo} & 9.14\,M  & 329.75 & 1.85 & 45.30 & 59.47  
        & 28.42 & 22.85 & 51.88 & 48.19 & 79.50 & 22.19 & 60.80 & 48.54\\
\rowcolor{myellow} EfficientViT-b1\cite{efficientvit} & 4.79\,M & 9.18 & 181.79 & 58.81 & 71.15  & 72.17 & 38.94 & 59.79 
        & 53.23 & 88.42 & 32.79 & 71.40 & 53.74\\
UniRepLKNet\cite{unireplknet} & 5.83\,M & 178.36 & 9.39 & 64.80 & 76.11   & \cellcolor{ll}{79.37} & 43.67 & 68.07 
        & 60.71 & 87.83 & 40.58 & 78.47 & 59.74\\    
\hline 
\multicolumn{3}{l}{\textit{{\color{Gray} light-weight medical network}}} &&&&& \\
UNeXt~\cite{unext}   & 1.47\,M   & 650.48 & 0.58 & 57.22 & 69.99  
          & 69.35 & 36.47 & 63.21 & 50.45 & 85.09 & 28.87 & 72.34 & 52.02\\
\rowcolor{myellow} MedT~\cite{medt}    & 1.37\,M   & 22.97  & 2.40 & 43.51 & 55.21   
          & 67.49 & \ \;1.63  & 61.82 & 49.81 & 36.11 & 20.26 & 66.33 & 44.64\\
TinyU-Net~\cite{tinyunet} & 0.48\,M & 359.50 & 1.66 & 59.31 & 71.81 & 69.95 & 42.24 & 67.18 & 58.99 & 84.63 & 31.66 & 71.73 & 48.50 \\

\hline
\rowcolor{myellow} \textbf{Mobile U-ViT}  & 1.39\,M   & 326.24 & 2.51 & \cellcolor{ll}{68.17} & \cellcolor{l}{79.13}  
             & \cellcolor{g}{\textbf{79.64}} & \cellcolor{ll}{45.96} & \cellcolor{l}{74.93} & \cellcolor{l}{68.69} & \cellcolor{g}{\textbf{90.40}} & \cellcolor{ll}{43.51} & \cellcolor{g}{\textbf{81.21}} & \cellcolor{ll}{60.98}\\
\rowcolor{myellow} \textbf{Mobile U-ViT-L}  & 7.88\,M  & 279.42  & 3.70 & \cellcolor{g}{\textbf{69.09}} & \cellcolor{g}{\textbf{79.90}} 
             & \cellcolor{l}{78.99} & \cellcolor{g}{\textbf{49.14}} & \cellcolor{ll}{72.55} & \cellcolor{ll}{68.29} & 88.87 & \cellcolor{g}{\textbf{50.10}} & 79.57 & \cellcolor{g}{\textbf{65.19}}\\
\hline
             
\end{tabular}
}
\end{table*}

\renewcommand{\multirowsetup}{\centering}  
\begin{table*}[h!]
\centering
\caption{Results on volume-level CT Datasets. Best results are highlighted as \colorbox{g}{\bf \!first\!}, \colorbox{l}{\!second\!} and \colorbox{ll}{\!third\!}. Note that GFLOPs and FPS are calculated for sub-volume crops (96 $\times$ 96 $\times$ 96). \label{Tab.3D}}
\resizebox{1\linewidth}{!}
{
\begin{tabular}{l | c c c | c c c | ccc | c c c}
\hline
\multirow{2}{*}{Network} & \multirow{2}{*}{Params↓} & \multirow{2}{*}{GFLOPs↓} & \multirow{2}{*}{FPS↑} & \multicolumn{3}{c|}{BTCV} & \multicolumn{3}{c|}{KiTS'23} & \multicolumn{3}{c}{Avg. Performance} \\
\cline{5-13}
& & & & Dice\%↑ & HD95↓ & ASD↓ &  Dice\%↑ & HD95↓ & ASD↓ &  Dice\%↑ & HD95↓ & ASD↓ \\
\cline{1-2}
\hline
UNETR (WACV'22)~\cite{unetr} & 92.61\,M & 82.70  & 52.00 & 71.67 & 10.92 & 3.32 & 56.91 & 28.11 & 12.96 & 64.29 & 19.51 & 8.14 \\ 
Swin UNETR (CVPR'22)~\cite{swinunetr} & 61.98\,M & 329.84 & 12.16 & 74.95 & 11.11 & 3.59 & 67.05 & 22.72 & 10.55 & 71.00 &  16.92 & 7.07 \\ 
MedNeXt (MICCAI'23)~\cite{mednext} & 10.51\,M & 74.58 & 19.19 & \cellcolor{ll}{76.14} & \cellcolor{l}{8.57} & \cellcolor{ll}{2.78} & \cellcolor{ll}{70.30} & \cellcolor{l}{13.51} & \cellcolor{g}{\textbf{4.27}} &  \cellcolor{ll}{73.22} & \cellcolor{l}{11.04} & \cellcolor{l}{3.53}  \\
SegMamba (MICCAI'24)~\cite{segmamba} & 65.18\,M & 659.04 & 9.63 & \cellcolor{l}{76.54} & 10.47 & 3.52 & \cellcolor{g}{\textbf{71.52}} & 17.42 & 6.51 &  \cellcolor{l}{74.03} & 13.95 & 5.02 \\

\hline
Mobile U-ViT (3D) & 2.75\,M & 28.56 & 55.13 & 76.01 & \cellcolor{ll}{8.67} & \cellcolor{l}{2.63} & 68.49 & \cellcolor{ll}{15.33} & \cellcolor{ll}{5.98} &  72.25 & \cellcolor{ll}{12.00} & \cellcolor{ll}{4.31} \\

Mobile U-ViT-L (3D) & 11.06\,M & 110.88 & 28.24 & \cellcolor{g}{\textbf{78.42}} & \cellcolor{g}{\textbf{6.56}} & \cellcolor{g}{\textbf{2.18}} & \cellcolor{l}{70.93} & \cellcolor{g}{\textbf{13.39}} & \cellcolor{l}{4.75} &  \cellcolor{g}{\textbf{74.68}} & \cellcolor{g}{\textbf{9.98}} & \cellcolor{g}{\textbf{3.47}} \\

\hline
\end{tabular}
}
\end{table*}

\begin{figure*}[t!]
  \centering
   \includegraphics[width=0.94\linewidth]{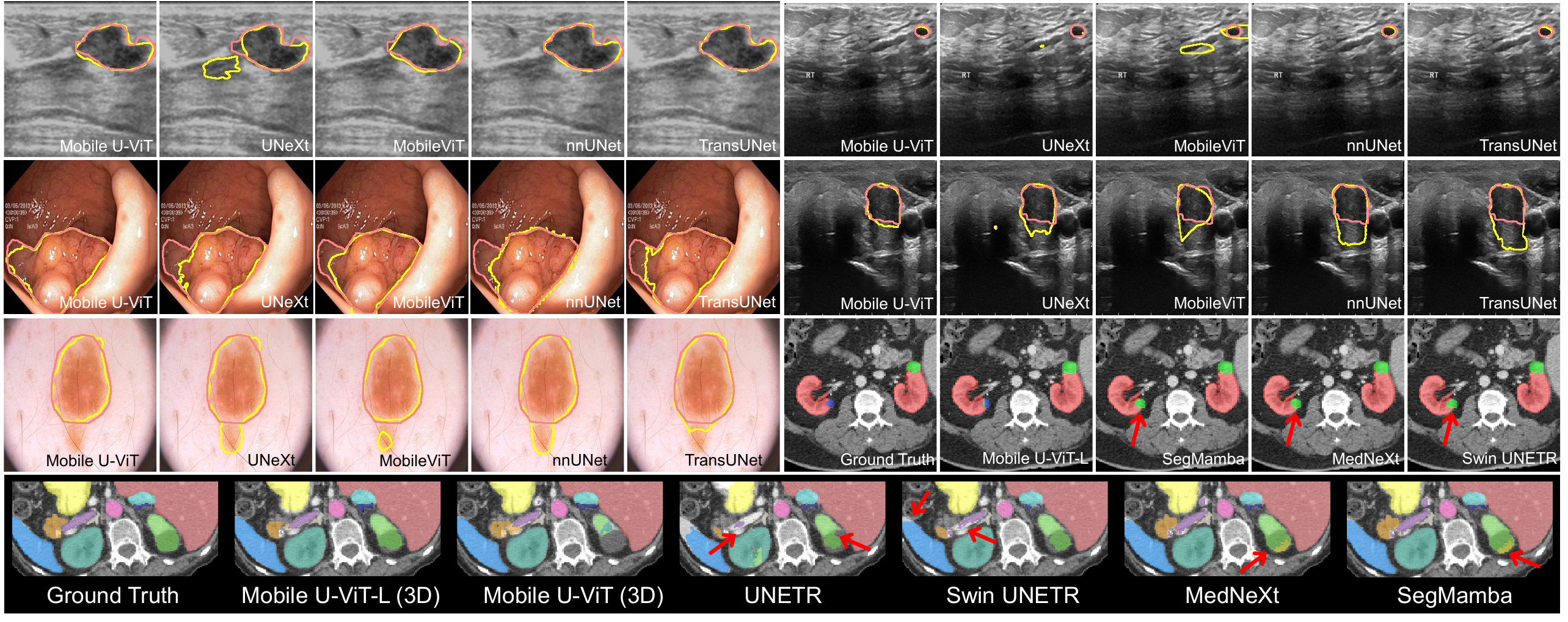}
   \caption{Visualization Results. Row 1 left - BUS samples, Row 1 right - BUSI samples, Row 2 left – Kvasir samples, Row 2 right - TNSCUI samples. Row 3 left - ISIC samples, Row 3 right - KiTS'23 samples.  Row 4 - BTCV samples. The \textcolor{yellow}{yellow line} is the prediction, and the \textcolor{pink}{pink line} is ground truth.}
      \label{fig:supply_other}
\end{figure*}

\begin{table*}[t!]
\centering
\fontsize{8}{10}\selectfont
\caption{Generalization experiment of lightweight model zero-shot on the unseen dataset. Source training dataset -> Zero-shot test dataset.   Best results are highlighted as \colorbox{g}{\bf \!first\!}, \colorbox{l}{\!second\!} and \colorbox{ll}{\!third\!}.} 
\label{tab:unseen}
{
\begin{tabular}{l|cc|cc|cc|cc|cc}
\hline
\multirow{3}{*}{Networks} & \multicolumn{4}{c|}{Ultrasound} & \multicolumn{2}{c|}{Dermoscopy} & \multicolumn{2}{c|}{Colonoscopy} & \multicolumn{2}{c}{\multirow{2}{*}{Avg}} \\
\cline{2 - 9} 
 &  \multicolumn{2}{c|}{BUSI -> BUS} & \multicolumn{2}{c|}{TUSCUI -> TUCC} & \multicolumn{2}{c|}{ISIC -> PH2} & \multicolumn{2}{c|}{Kvasir -> CVC-300} \\
\cline{2 - 11} 
 &  \multicolumn{1}{c}{IoU(\%)} & \multicolumn{1}{c|}{F1(\%)} & \multicolumn{1}{c}{IoU(\%)} & \multicolumn{1}{c|}{F1(\%)}& \multicolumn{1}{c}{IoU(\%)} & \multicolumn{1}{c|}{F1(\%)} & \multicolumn{1}{c}{IoU(\%)} & \multicolumn{1}{c|}{F1(\%)} & \multicolumn{1}{c}{IoU(\%)} & \multicolumn{1}{c}{F1(\%)} \\ \hline

UNeXt (MICCAI'22)~\cite{unext}    & 69.02  & 79.27 &54.83&66.04 & 82.00  & 89.57 & 65.75  & 75.23 & 67.90 & 77.53 \\ 
ERDUnet (TCSVT'23)~\cite{erdunet}	         & \cellcolor{ll}{71.75}	&80.72&59.66&70.06	&82.90	&90.06	&72.23	& 80.62 & \cellcolor{ll}{71.64} & 80.37 \\
RepViT-m3 (CVPR'24)~\cite{repvit}  & 65.07  & 76.38&57.28&68.88 & 79.95  & 88.27 &62.96  & 73.68 & 66.32 & 76.80 \\ 
MobileViT-s (ICLR'22)~\cite{mobilevit} &70.70  & \cellcolor{ll}{81.14} & \cellcolor{ll}{59.83} & \cellcolor{ll}{70.78} & 83.19  &90.42  & \cellcolor{ll}{72.50}  & \cellcolor{l}{81.92} & 71.56 & \cellcolor{ll}{81.07} \\
UniRepLKNet (CVPR'24)~\cite{unireplknet} & 66.19  & 75.65&55.53& 67.12 & \cellcolor{ll}{83.83}  & \cellcolor{ll}{90.80}  & 68.31  & 75.59  & 68.47 & 77.29   \\ 
TinyU-net (MICCAI'24)~\cite{tinyunet}    & 64.49  & 73.70 &57.76&69.03 & 81.96  & 89.53 & 71.22  & 78.50 & 68.86 &  77.69 \\ 
\hline
Mobile U-ViT (3D) (Ours)    & \cellcolor{l}{76.64} & \cellcolor{l}{85.35} & \cellcolor{l}{60.33} & \cellcolor{l}{71.11} & \cellcolor{l}{84.63} & \cellcolor{l}{91.05} & \cellcolor{l}{73.62} & \cellcolor{g}{\textbf{81.94}} & \cellcolor{l}{73.81} & \cellcolor{l}{82.36}   \\ 
Mobile U-ViT-L (3D) (Ours)   & \cellcolor{g}{\textbf{78.19}} & \cellcolor{g}{\textbf{86.45}} & \cellcolor{g}{\textbf{60.71}} & \cellcolor{g}{\textbf{71.44}}  & \cellcolor{g}{\textbf{85.25}} & \cellcolor{g}{\textbf{91.53}} & \cellcolor{g}{\textbf{73.65}} & \cellcolor{ll}{81.79} & \cellcolor{g}{\textbf{74.45}} & \cellcolor{g}{\textbf{82.80}}    \\ 
\hline
\end{tabular}
}
\end{table*}

\subsection{Experimental Results}

\subsubsection{2D results with different modalities.}
As shown in Table~\ref{tab:ExOnUltrasound}, for ultrasound (BUS, BUSI, TNSCUI), dermoscopy (ISIC), and polypscopy (Kvasir) images, 
our network Mobile U-ViT achieves superior results while maintaining a significantly smaller model size (1.39 M \textit{v.s.} 26.10 M) and improved computational efficiency (2.51 GFLOPs \textit{v.s.} 12.67 GFLOPs) compared to competitive nnUNet. Extensive experiments across ultrasound, dermoscopy, and polypscopy imaging modalities reveal that Mobile U-ViT-L achieves state-of-the-art segmentation performance, with IoU scores of 87.63\% (BUS), 73.91\% (BUSI), 83.31\% (ISIC), and 89.07\% (Kvasir)—surpassing nnUNet by 0.1\%, 1.8\%, 0.01\%, and 3.4\%, respectively. Even on the largest TNSCUI dataset, our method delivers competitive accuracy while remaining lighter and faster than existing approaches.

\textit{Mobile U-ViT's medical-targeted design proves effective.} 
Visualization Fig shows a significant domain gap between different medical imaging modalities. Existing lightweight CNNs exhibit limitations in long-range dependency modeling  (\textit{e.g.}, UNeXt in BUS), and hybrid architectures (\textit{e.g.}, TransUnet, UCTransNet) face challenges in balancing performance and computational efficiency. Mobile U-ViT addresses the domain gap and model design issues. Its modern, lightweight architecture delivers optimal performance across various modalities, with the Mobile U-ViT and Mobile U-ViT-L variant average IoU of 81.67\% and 82.43\% outperforming other recent heavy- and light-weight medical network. We visualize some examples in Fig.~\ref{fig:supply_other}. Mobile U-ViT effectively captures finer details and better aligns with the ground truth, showing fewer mismatches and more precise delineation of structures.

\textit{Mobile U-ViT excels not only in performance but also in computational efficiency.} As shown in Fig.~\ref{fig:performance}, it has significantly fewer parameters than heavy-weight hybrid medical networks like TransUnet and UCTransNet (105.32 \'M and 66.24 \'M \textit{v.s.} 1.39\'M). With reduced GFLOPs, Mobile U-ViT is highly efficient for resource-constrained environments. Additionally, its high FPS ensure fast inference, which addressing the requirements for real-time computer-aided clinical diagnostics across multiple imaging modalities. These features make Mobile U-ViT an ideal solution, offering high performance with low computational overhead for mobile medical applications.

\subsubsection{CT results in slice- and volume-level.} We further conduct experiments in both slice-level (Synapse) and volume-level (BTCV and KiTS'23) datasets, as shown in Tab.~\ref{tab:ct} and Tab.~\ref{Tab.3D}, respectively. 

\begin{figure}[t!]
  \centering
   \includegraphics[width=\linewidth]{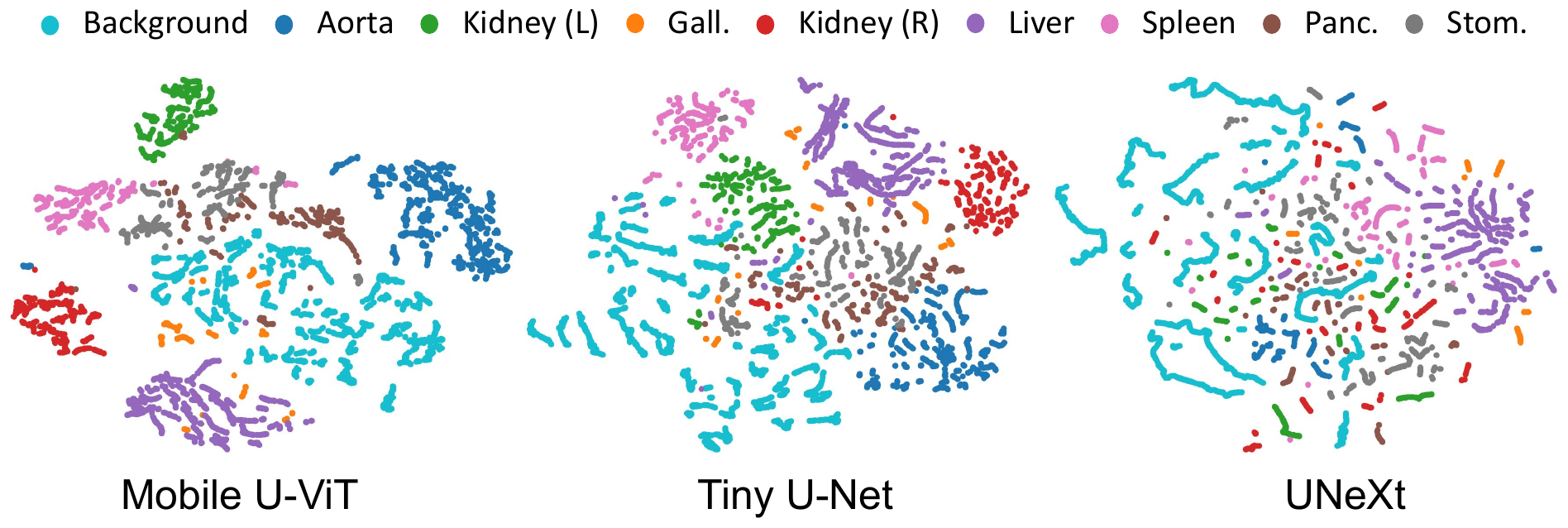}
   \vspace{-5mm}
   \caption{t-SNE visualization of encoder features extracted from Mobile U-ViT, TinyU-Net, and UNeXt on Synapse.}
      \label{fig:tsne}
      \vspace{-3mm}
\end{figure}

\begin{figure}[t!]
  \centering
   \includegraphics[width=0.88\linewidth]{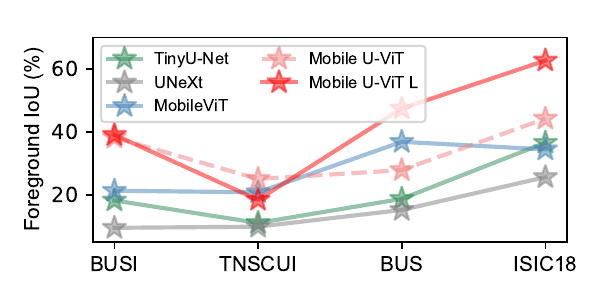}
   \vspace{-5mm}
   \caption{Foreground capture capability is analyzed by foreground-background eigenvalue ratio. Our method demonstrates generally improved performance.}
      \label{fig:foreground}
\end{figure}

Slice-level evaluation in Tab.~\ref{tab:ct} shows that contemporary lightweight networks experience significant performance drops compared to heavy-weight networks. We analyze the feature space by t-SNE, as shown in Fig.~\ref{fig:tsne}. The result reveals a pronounced entanglement among classes of lightweight networks employing channel reduction (\textit{e.g.} Tiny U-Net, UNeXt), suggesting the suboptimal discriminative capacity of their learned feature representations. In contrast, Mobile U-ViT is able to maintain the feature discrimination among classes. This is also validated by the quantitative results shown in Tab.~\ref{tab:ct} and qualitative results shown in Fig.~\ref{fig:supply_other}. Mobile U-ViT achieves the best segmentation accuracy (Dice of 79.90\%), while identifying structures and boundaries with minimal error. Notably, its compact size and high FPS further confirm its suitability for real-time medical image segmentation on edge devices.

\begin{figure}[t!]
  \centering
   \includegraphics[width=0.94\linewidth]{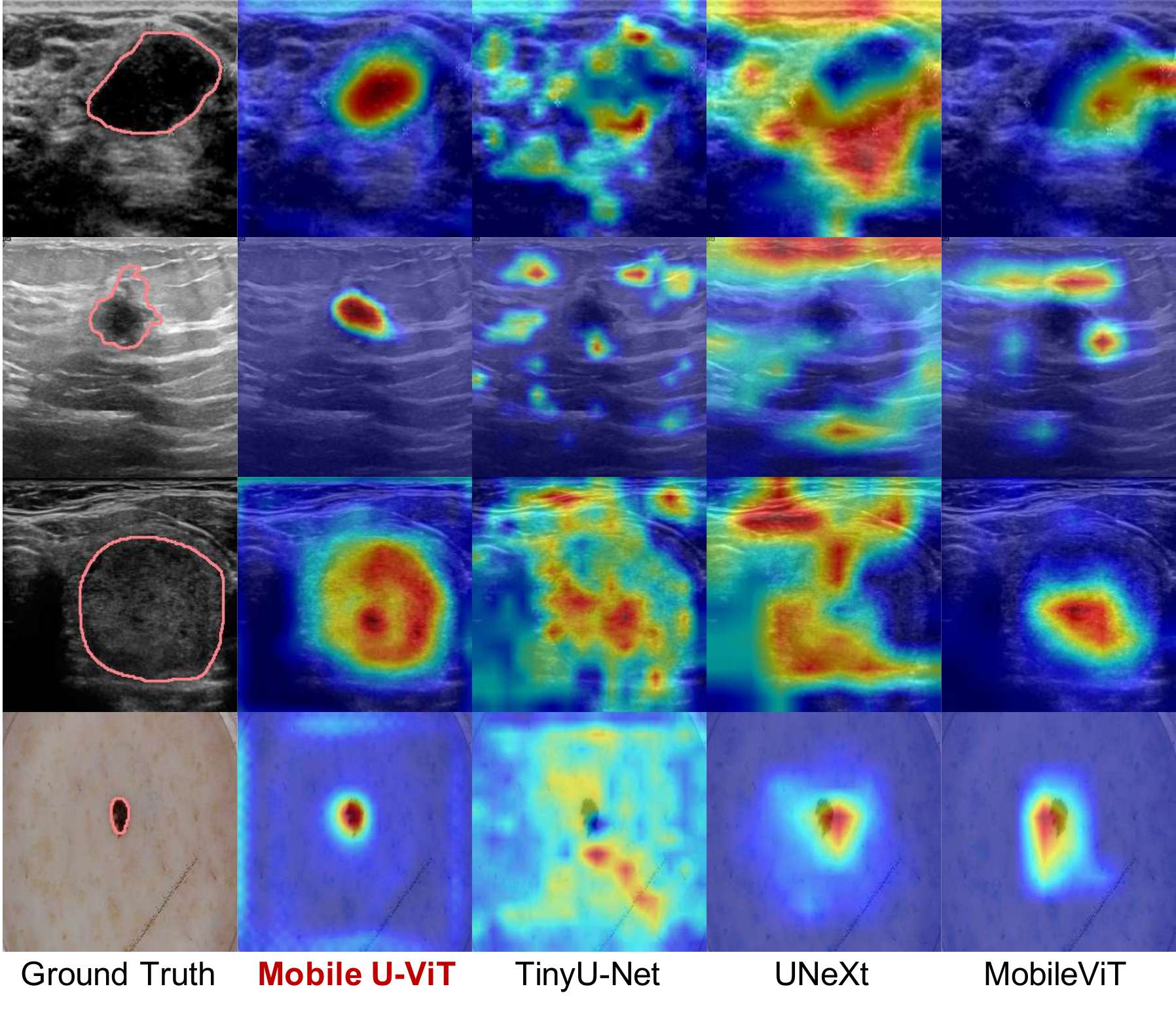}
   \vspace{-3mm}
   \caption{Visualization of Grad-CAM activations corresponding to the encoder outputs.}
      \label{fig:cam}
      \vspace{-6mm}
\end{figure}

As shown in Tab.~\ref{Tab.3D},
we extend Mobile U-ViT to ``3D'' and evaluate it on the BTCV and KiTS'23 datasets. While Mamba-based models like SegMamba have gained attention in medical segmentation, our 3D Mobile U-ViT-L outperforms SegMamba, improving average Dice by 0.65\% (74.68\% \textit{v.s.} 74.03\%) with significantly lower computational costs (11.06 M vs. 65.18 M). 
Compared to other 3D networks, Mobile U-ViT achieves SOTA performance while maintaining lower computational overhead. The  results of each segmentation target are shown in Fig~\ref{fig:teaser} (b). Even on the largest 3D dataset KiTS'23, our method delivers competitive accuracy in HD95 and ASD, while remaining lighter and faster than existing approaches. We attribute the improvement to the strong local and global representation abilities acquired from our design, enabling Mobile U-ViT to segment organs and tumor with different scales. 

\begin{table*}[!h]
\caption{Ablation study on each block using Synapse.
MV2: MobileNetV2 block as patch embedding; MViT: MobileViT block~\cite{mobilevit} as bottom encoder;  MV2 (LK): MobileNetV2 block~\cite{mobilenetv2} with large kernels; skip1: skip-connections with the top stage of encoder; skip2: skip-connections with the top and second stage of encoder; skip3: skip-connections with whole encoder.\label{tab:ablation}}
\resizebox{0.9\linewidth}{!}
{
\begin{tabular}{l|l c c | r r r r}
\hline 

\multirow{3}{*}{\footnotesize{Line\#}}  & \multicolumn{3}{c|}{Network}
& \multicolumn{4}{c}{Metrics(\%)}\\
\cline{2-8}
& \multirow{2}{*}{Patch embedding + Transformer backbone} & Skip- & Cascaded & \multirow{2}{*}{Params↓}  & \multirow{2}{*}{GFLOPs↓} &  \multirow{2}{*}{FPS↑} & \multirow{2}{*}{Jaccard↑}\\
& & Connection & Decoder & &&&\\
\hline
\rowcolor{myellow} 1. & \textbf{MobileViT} (MV2 + MViT) &  &  & 10.66\,M & 2.13 &  291.44 & 40.84\\
2.& MV2 (LK) + MViT  &  &  &  10.69\,M & 2.22 & 301.04 & 41.16\\
3.& ResNet34 + MViT & & & 53.05\,M & 22.56 & 141.95 & 41.98 \\
4.& ConvUtr + MViT &   &  & 10.64\,M & 7.69 & 236.68 & 56.98\\
5.& MV2 + MViT &  & \checkmark & 5.91\,M & 4.01 & 313.85 & 46.09\\
6.& MV2 + MViT & \checkmark & \checkmark & 6.18\,M & 4.44 & 306.86 & 65.56\\
7. & MV2 + LKLGL + MViT &  &  & 10.50\,M & 1.91 & 284.02 & 41.10\\
8.& ConvUtr + MViT & \checkmark & \checkmark & 4.61\,M & 4.06 & 277.86 & 65.56 \\
9.& ConvUtr + LKLGL + MViT &  &  & 10.46\,M & 6.70 & 244.85 & 54.90\\
\rowcolor{myellow} 10.& ConvUtr + LKLGL + MViT & \checkmark & \checkmark & 4.42\,M & 3.84 & 270.97 &  \textbf{68.90} \\
\hline
11.& ResNet34 + ViT  &  & \checkmark & 25.65\,M & 85.95 & 77.75 & 63.76\\
12.&ResNet34 + LKLGL  &  & \checkmark & 22.11\,M & 81.41 & 100.06 & 63.25\\
13.&ConvUtr + LKLGL + ViT &  &   &  2.61\,M & 3.25 & 350.45 & 54.18 \\
14.&ConvUtr + LKLGL + ViT &  & \checkmark & 1.34\,M & 2.39 & 339.16 & 64.40 \\
15.&ConvUtr + LKLGL + ViT   & skip1  & \checkmark  & 1.34\,M & 2.43 & 332.45 & 67.02\\

16.&ConvUtr + LKLGL + ViT   & skip2  & \checkmark  & 1.35\,M & 2.46 & 328.16 & 67.51\\
\rowcolor{myellow} 17.&\textbf{Mobile U-ViT} (ConvUtr + LKLGL + ViT)   & skip3  & \checkmark  & 1.39\,M & 2.50 & 322.72 & \textbf{68.17}\\
\hline
18.&ConvUtr + ViT  & horizontal skip3 & \checkmark & 1.67\,M & 3.34 & 391.08 & 66.50\\
19.&ConvUtr + ViT  & skip3 & \checkmark &  1.76\,M & 2.91 & 416.22 & 67.26\\

\hline
20.&Mobile U-ViT (convolution downsampling) & skip3  & \checkmark  & 1.40\,M & 2.52 & 316.85 & 66.29 \\
21.& Mobile U-ViT (maxpooling downsampling) & skip3  & \checkmark  & 1.39\,M & 2.50 & 322.72 & 68.17\\
\hline
\end{tabular}

}
\end{table*}

\subsection{Analysis}
\subsubsection{Generalization Analysis}

To evaluate the domain generalizability of Mobile U-ViT trained in each modality, we test model on
external dataset from unseen clinical settings. 
The results presented in Tab.~\ref{tab:unseen} highlight the superiority of Mobile U-ViT's robustness across zero-shot segmentation tasks in various image modalities. Our model consistently outperforms contemporary lightweight models, achieving the highest performance with significant improvement of 2.17\% and 2.81\% on IoU, respectively. Considering the diversity of clinical scenarios, its lightweight design and robustness further enhances its applicability in real-world clinical settings.


\subsubsection{Statistic Analysis of Encoded Feature}

To evaluate the effectiveness of the encoded features, we measure the accuracy of feature concentration by calculating the IoU between the ground truth segmentation masks and the highlighted regions (threshold by 0.4) in the encoder output. The results in Fig.~\ref{fig:foreground} offer a quantitative assessment of how well the model focuses on the relevant foreground regions. Features from our Mobile U-ViT (shown in red lines) generally gains the best IoU score across different datasets, validating the effectiveness of our proposed encoder design. 

Additionally, as shown in Fig.~\ref{fig:cam} the Grad-CAM visualizations reveal that our Mobile U-ViT effectively captures both global contextual information and fine-grained local details. Compared to other methods, the activation maps generated by our encoder exhibit broader coverage of the foreground regions while maintaining sharp focus on object boundaries, demonstrating the model's ability to balance semantic understanding with precise localization. More results can be found in Appendix.

\subsubsection{Evaluation on NVIDIA Jetson Nano}
Edge medical devices typically adopt diverse processing architectures. We evaluate our model on the entry-level NVIDIA Jetson Nano (472 GFLOPS), which is widely used in handheld ultrasound systems (e.g., Butterfly iQ/iQ+, Clarius). Medical diagnostic applications require both high accuracy and real-time preview. As shown in Tab.~\ref{tab:eval_device}, our models achieve SOTA performance across multiple datasets, with actual frame rates of 51.22 FPS (Mobile U-ViT) and 19.52 FPS (Mobile U-ViT-L), satisfying clinical deployment needs. INT8 quantization further improves memory and computational efficiency, making our approach well-suited for edge medical devices.

\renewcommand{\multirowsetup}{\centering}  
\begin{table}[t!]

\caption{Actual Edge devices Evaluation. \label{tab:eval_device}}
{
\begin{tabular}{c c c | c}
\hline 
\multicolumn{2}{c}{Inference time (ms)} & \multirow{2}{*}{Memory (MB)} & \multirow{2}{*}{Jaccard (\%)} \\
\cline{1-2}
Theoretical & Actual &  &  \\

\hline
Mobile U-ViT & 5.34 & 19.52 & 17.58 \\
Mobile U-ViT-L & 22.19 & 37.43 & 23.52 \\
\hline
\end{tabular}
}
\end{table}

\subsection{Ablations}
\subsubsection{Component Ablation Study}

As shown in Tab.~\ref{tab:ablation}, we set MobileViT as the backbone in Line 1 and progressively integrate our proposed modules to evaluate the effect of each component.

Firstly, comparing Line 2-4, we find that ConvUtr demonstrates a substantial impact, achieving a Jaccard index of 56.98\% compared to 41.16\% without ConvUtr (an improvement of 36\%). This improvement can be attributed to ConvUtr’s larger receptive field and its ability to capture richer representations. Additionally, increasing the kernel size in the patch embedding module of MobileViT also yields performance improvements (Lines 1 and 2), but adding the model depth (Line 4) results in performance degradation.

Moreover, it is noteworthy that a naive application of LKLGL does not yield performance gains (Lines 2 and 7), but demonstrates efficiency improvements (also seen in Lines 11 and 12). When LKLGL is integrated into ConvUtr encoder (Line 7-10), performance is significantly enhanced, indicating that ConvUtr can boost synergy in CNN-Transformer hybrid structure. We attribute it to ConvUtr's ability to emulate the learning patterns of Transformers. This synergy facilitates the interaction between global and local information flows inherent to LKLGL (Lines 8 and 10).
It is also worth-noting that, due to the large parameter number of MViT, we replace it with ViT (1.39M \textit{v.s.} 4.42M parameters, representing a 68.55\% reduction). Although this substitution leads to a slight performance drop, the substantial reduction in model complexity makes it a favorable trade-off.

Furthermore, we investigate the impact of cascaded decoders on performance and observe that all methods incorporating cascaded decoders yielded notable improvements (see ``Cascaded Decoder'' column with $\checkmark$). This can be attributed to the strengths of the U-shaped designs, which is particularly effective in medical imaging tasks by enabling accurate dense predictions in medical images with blurred boundaries and high noise levels. Additionally, comparing Line 19 with Line 18, our strategy of using downsampled skip connections in Line 19 is effective and efficient (Jaccard of 67.26\% \textit{v.s.} 66.50\% and GFLOPs of 2.91 \textit{v.s.} 3.34).

Finally, we examine the impact of different downsampling strategies on medical image feature extraction. Specifically, we replace all downsampling operations with convolutional downsampling (Lines 20 and 21). The results indicate a decline in segmentation performance under this modification, further highlighting the critical role of max pooling in effectively extracting features from sparse and noisy medical images.

\subsubsection{Kernel Ablation Study}
Similar to prior works (e.g., ConvNeXt~\cite{liu2022convnet}, CMUNeXt~\cite{tang2024cmunext}), we define kernels larger than 3 as large kernels. Unlike methods applying large kernels throughout the encoder (e.g., ConvNeXt~\cite{liu2022convnet}), our design emphasizes preserving fine-grained semantics by smoothly expanding receptive fields. As validated by a addition experiment of Tab.~\ref{tab:ablation} (Line 17) with kernel size ConvUtr (K1,K2,K3) and LKLGL (K4) of method changed, abrupt transitions (3$\rightarrow$9 or 7$\rightarrow$9) degrade performance, while gradual transitions (3$\rightarrow$7$\rightarrow$9) yield better results.

\renewcommand{\multirowsetup}{\centering}  
\begin{table}[h!]

\caption{Ablation on Kernel. \label{tab:ablation_kernel}}
{
\begin{tabular}{c c c c c}
\hline 

K1 & K2 & K3 & K4 & Jaccard (\%) \\
\hline
3 &	3 &	3 & 9 & 64.27 \\
7 &	7 &	7 & 9 & 67.15 \\
3 &	3 &	7 & 9 & 68.17 \\
\hline
\end{tabular}
}
\end{table}

\subsubsection{LKLGL Ablation Study}
LKLGL addresses medical imaging properties such as sparse local information, blurred boundaries, and high noise levels. It combines large-kernel depthwise convolution for local aggregation and sparse global attention to reduce complexity while preserving long-range dependencies. After global sparse attention, local propagation via transposed convolution redistributes global context back to fine-grained detailed, refining local information that are critical for segmentation. While ablation is conducted on Synapse, LKLGL remains effective across modalities (\textit{e.g.}, ultrasound, dermoscopy) that share challenges like low contrast and boundary ambiguity (shown in Fig.~\ref{fig:teaser}). We also perform additional ablation studies on other modalities to evaluate the effectiveness of LKLGL. As shown in Tab.~\ref{tab:ablation_lklgl}, applying LKLGL (w/) brings consistent gains of 0.59, 0.75, 0.75, 0.52, and 0.43 on BUS, BUSI, TNSCUI, ISIC, and Kvasir, confirming its generalizability across medical domains.

\renewcommand{\multirowsetup}{\centering}  
\begin{table}[h!]

\caption{Ablation on LKLGL. \label{tab:ablation_lklgl}}
{
\begin{tabular}{l | c c c c c}
\hline 
\multirow{2}{*}{LKLGL} & \multicolumn{5}{c}{IoU (\%)} \\
& BUS & BUSI & TNSCUI & ISIC & Kvasir \\
\hline
\textit{w/o} & 86.88 & 72.29 & 76.95 & 82.71 & 86.66 \\
\textit{w/} & 87.28 & 72.88 & 77.70 & 83.23 & 87.28 \\
\hline
\end{tabular}
}
\end{table}

\section{Conclusion}
\label{sec:Conclusion}
In this work, we present Mobile U-ViT, a novel hybrid lightweight network designed to address the challenges of mobile medical image analysis. By leveraging a combination of ConvUtr—a lightweight, Transformer-inspired CNN patch embedding—and LKLGL modules, Mobile U-ViT effectively bridges the gap between computational efficiency and performance in medical image segmentation. Additionally, the cascaded decoder with downsampled skip-connections ensures efficient integration of local and global features, making the model ideal for resource-constrained environments. Through extensive experiments, Mobile U-ViT has demonstrated state-of-the-art performance across a wide range of 2D and 3D datasets, showcasing its zero-shot generalization capability and proving its robustness in diverse medical imaging tasks. Our work excels in maintaining low computational complexity, a reduced parameter count, and a high real-time frame rate, all while preserving or even improving accuracy in general medical segmentation tasks. This innovative design successfully balances the lightweight nature of ViT with robust performance. We hope our approach will inspire further advancements in efficient and high-performing models for mobile medical image analysis.

\begin{acks}
Supported by Natural Science Foundation of China under Grant
62271465, 62376153, 62402318, 24Z990200676, Suzhou Basic
Research Program under Grant SYG202338, and Joint Funds of the National Natural Science Foundation of China (U22A2033).
\end{acks}

\clearpage
\appendix
\section*{Appendix}
\addcontentsline{toc}{section}{Appendix}

\vspace{1em}

\section{Datasets}

\subsection{Training datasets}

\textbf{BUS}. The Breast UltraSound (BUS) dataset~\cite{DATA_BUS} contains 562 breast ultrasound images collected using five different ultrasound devices, including 306 benign cases and 256 malignant cases, each with corresponding ground truth.

\noindent\textbf{BUSI}. The Breast UltraSound Images (BUSI) dataset~\cite{DATA_BUSI} collected from 600 female patients includes 780 breast ultrasound images, covering 133 normal cases, 487 benign cases, and 210 malignant cases, each with corresponding ground truth. Following recent studies \cite{unext, cmunet}, we only utilize benign and malignant cases from this dataset.

\noindent\textbf{TNSCUI}. The Thyroid Nodule Segmentation and Classification in Ultrasound Images 2020 (TNSCUI) dataset\footnote{\url{https://tn-scui2020.grand-challenge.org/Dataset/}} is collected by the Chinese Artificial Intelligence Alliance for Thyroid and Breast Ultrasound (CAAU). It includes 3644 cases of different ages and genders, each with corresponding ground truth.
    
\noindent\textbf{ISIC 2018}. The International Skin Imaging Collaboration (ISIC 2018) dataset~\cite{DATA_ISIC2018} contains 2,594 dermoscopic lesion segmentation images, each with corresponding ground truth.

\noindent\textbf{Kvasir}. The Kvasir dataset~\cite{kvasir} contains images of the gastrointestinal (GI) tract and gathered using the apparatus called colonoscope at Vestre Viken Health Trust (VV) situated in Norway. It contains 900 training images and 100 testing images in the dataset. As pixel-by-pixel segmentations, the ground truth annotations for each image are included in the dataset.

\noindent\textbf{Synapse}. Synapse is a CT multi-organ segmentation dataset from the MICCAI 2015 Multi-Atlas Abdomen Labeling Challenge\footnote{\url{https://tn-scui2020.grand-challenge.org/Dataset/}}. It comprises abdominal CT scans of 8 organs from 30 cases (3779 axial images). Each CT volume consists of $85 \sim 198$ slices of 512$\times$512 pixels, with a voxel spatial resolution of ($[0.54 \sim 0.54]\times[0.98 \sim 0.98]\times[2.5 \sim 5.0]$) mm$^3$.

\noindent\textbf{BTCV}. The BTCV dataset~\cite{btcv} consists of 30 subjects with abdominal CT scans where 13 organs are annotated by interpreters under supervision of clinical radiologists at Vanderbilt University Medical Center.

\noindent\textbf{KiTS'23}. The 2023 Kidney and Kidney Tumor Segmentation challenge (KiTS'23) dataset~{\cite{kits23}} consists of 489 CT scans for the Kidney and its Tumor and Cyst annotations from M Health Fairview medical center.

\begin{table}[!t]

\centering
\caption{Overview of implement training details.\label{tab:settings}}
\resizebox{0.94\linewidth}{!}
{
\begin{tabular}{l l}
\hline 
\multicolumn{2}{l}{\textit{{\color{Gray} 2D medical image pre-processing}}} \\
\hline
image size & $256 \times 256$ \\
Augmentation & Random Rotate, Flip, Normalize \\
\hline
\multicolumn{2}{l}{\textit{{\color{Gray} 2D training settings}}} \\
\hline
Training epoch & 300 \\
Optimizer & SGD \\
Weight decay & 1e-4 \\
Optimizer momentum & 0.9 \\ 
Optimizer LR & 1e-2 \\
Batch size & 8 \\
LR schedule & warmup cosine \\
\hline
\multicolumn{2}{l}{\textit{{\color{Gray} 3D medical volume pre-processing}}} \\
\hline
Spacing & $1.5 \times 1.5 \times 2.0~(mm)$ \\
Intensity & $[-175, 250]$ \\
Sub-volume size & $96 \times 96 \times 96$ \\
Sub-crops & 4 \\
Augmentation & Random Rotate, Flip, Scale, Shift \\
\hline
\multicolumn{2}{l}{\textit{{\color{Gray} 3D training settings}}} \\
\hline
Optimizer & AdamW \\
Optimizer LR & 1e-4 \\
Weight decay & 1e-5 \\
Batch size & $1 \times 4 = 4$ \\
Swin batch size & 1 \\
Inference & sliding window \\
\hline
\end{tabular}
}
\end{table}

\subsection{Zero-shot datasets}

\noindent\textbf{TUCC}. The Thyroid Ultrasound Cine-clip (TUCC) dataset\footnote{\url{https://aimi.stanford.edu/datasets/thyroid-ultrasound-cine-clip}} is collected data from 167 patients with biopsy-confirmed thyroid nodules (n=192) at the Stanford University Medical Center. The dataset consists of ultrasound cine-clip images, radiologist-annotated segmentations, patient demographics, lesion size and location, TI-RADS descriptors, and histopathological diagnoses. We test its 17K video frame.

\noindent\textbf{PH2}. The PH2 dataset~\cite{ph2} contains 200 dermoscopic images. The manual segmentation, clinical diagnosis, and identification of many dermoscopic structures, carried out by skilled dermatologists, are included in the PH2.

\noindent\textbf{CVC-300}. CVC-300~\cite{cvc300} is selected from the EndoScene test dataset, which contains 912 images from 44
colonoscopy sequences. Following~\cite{cvc300son}, CVC-300 was used as a test dataset with 60 test samples.

\renewcommand{\multirowsetup}{\centering}  
\begin{table*}[t!]

\caption{Results on Ultrasound Datasets ($\text{Mean}_{\text{Std}}$).  Best results are highlighted as \colorbox{g}{\bf \!first\!}, \colorbox{l}{\!second\!} and \colorbox{ll}{\!third\!}. \colorbox{myellow}{Gray} background indicate hybrid architectural method.  \label{tab:ExOnUltrasound}}
\resizebox{0.94\linewidth}{!}
{
\begin{tabular}{l rrr|cc|cc|cc}
\hline 

\multirow{3}{*}{Network} & \multirow{3}{*}{Params↓} & \multirow{3}{*}{FPS↑} & \multirow{3}{*}{GFLOPs↓} & \multicolumn{6}{c}{Ultrasound}  \\
\cline{5-10}
&&&& \multicolumn{2}{c|}{BUS  (\%)} & \multicolumn{2}{c|}{BUSI  (\%)} & \multicolumn{2}{c}{TNSCUI  (\%)} \\
\cline{5-10}
&&&& IOU &F1 & IOU &F1 & IOU &F1\\
\hline
\multicolumn{3}{l}{\textit{{\color{Gray} heavy-weight medical network}}} &&&&& \\

\rowcolor{mygray} U-Net (MICCAI'15)~\cite{unet} & 34.52\,M & 139.32 & 65.52 & $86.73_{1.41}$ & $92.46_{1.17}$ & $68.61_{2.86}$ & $76.97_{3.10}$
                                 & $75.88_{0.18}$ & $84.24_{0.07}$ \\
\rowcolor{mygray} CMU-Net (ISBI'23)~\cite{cmunet} & 49.93\,M & 93.19  & 91.25 & $87.18_{0.59}$ & $92.89_{0.41}$ & $71.42_{2.65}$ & $79.49_{2.92}$ 
                                 & $77.12_{0.49}$ & $85.35_{0.50}$ \\
\rowcolor{mygray} nnUNet (NM'21)~\cite{nnunet}  & 26.10\,M   & ---    & 12.67   & \cellcolor{l}{$87.51_{1.01}$} & \cellcolor{l}{$93.02_{0.73}$} & \cellcolor{ll}{$72.11_{3.51}$} & \cellcolor{ll}{$80.09_{3.77}$} & \cellcolor{g}{\textbf{78.99}$_{0.14}$} & \cellcolor{g}{\textbf{86.85}$_{0.15}$}\\
Swin-Unet (ECCV'22)~\cite{swinunet} & 27.14\,M  & 392.21 & 5.91  & $85.27_{1.24}$ & $91.99_{0.75}$ & $63.59_{4.96}$ & $76.94_{4.12}$ 
                                    & $75.77_{1.29}$ & $85.82_{0.91}$\\ 
\rowcolor{myellow} TransUnet (ArXiv'21)~\cite{transunet} & 105.32\,M & 112.95 & 38.52 & \cellcolor{ll}{$87.35_{1.24}$} & $92.88_{0.88}$ & $71.39_{2.37}$ & $79.85_{2.59}$ 
                                    & $77.63_{0.14}$ & $85.76_{0.20}$\\  
\rowcolor{myellow} UCTransNet (AAAI'22)~\cite{uctransnet} & 66.24\,M & 57.11 & 32.98 & $85.98_{1.34}$ & $92.03_{1.05}$
& $68.47_{2.61}$ & $77.16_{2.90}$ & $73.87_{0.24}$ & $82.63_{0.08}$\\
 MissFormer (TMI'22)~\cite{missformer} & 35.45\,M & 151.46 & 7.25 & $81.92_{1.20}$ & $89.46_{0.92}$ 
& $63.29_{2.99}$ & $73.47_{3.11}$ & $68.26_{2.47}$ & $76.71_{0.16}$\\
                                    
\hline                                    
\multicolumn{3}{l}{\textit{{\color{Gray} light-weight natural network}}} &&&&& \\

\rowcolor{myellow} MobileViT-s (ICLR'21)~\cite{mobilevit} & 10.66\,M  & 301.11 & 2.13  & $82.57_{1.38}$ & $89.99_{1.17}$ & $64.28_{3.78}$ & $74.68_{3.81}$ 
                                    & $71.64_{0.28}$ & $81.60_{0.25}$ \\
\rowcolor{myellow} EdgeViT-s (ECCV'22)~\cite{edgevits} & 16.49\,M  & 291.44 & 2.71  & $81.32_{1.23}$ & $89.13_{0.99}$ & $61.12_{3.69}$ & $71.79_{4.00}$ 
                                    & $68.74_{0.29}$ & $79.19_{0.36}$\\

\rowcolor{myellow} EMO-6m (ICCV'23)~\cite{edgevits} & 9.14\,M  & 329.75 & 1.85  & $84.89_{0.86}$ & $91.58_{0.63}$ & $67.50_{4.26}$ &$77.71_{4.23}$ & $74.02_{0.39}$ & $83.53_{0.25}$  \\ 
\rowcolor{myellow} EfficientViT-b1 (ICCV'23)~\cite{efficientvit}  & 4.79\,M & 181.79 & 9.18 & $84.81_{0.97}$ & $91.36_{0.70}$ & $67.21_{2.72}$ & $76.77_{3.03}$ & $71.18_{4.92}$ & $80.73_{4.02}$  \\
\rowcolor{myellow} RepViT-m3 (CVPR'24)~\cite{repvit}   & 14.37\,M  & 238.90 & 2.79   & $76.51_{1.38}$ & $85.79_{1.12}$ & $56.07_{3.51}$ & $67.62_{3.98}$ & $66.21_{0.66}$ & $77.09_{0.49}$ \\
UniRepLKNet (CVPR'24)~\cite{unireplknet}  &5.83\,M & 178.36 & 9.39 & $84.13_{1.32}$ &$90.76_{1.01}$ 
&$ 65.26_{1.72}$ & $73.96_{1.99}$ & $67.73_{0.92}$ & $77.20_{0.90}$ \\

\multicolumn{3}{l}{\textit{{\color{Gray} light-weight medical network}}} &&&&& \\

\rowcolor{myellow} MedT (MICCAI'21)~\cite{medt}  & 1.37\,M   & 22.97  & 2.40  & $80.81_{2.77}$ & $88.78_{1.96}$ & $63.36_{1.56}$ & $73.37±1.63$ 
                                    & $71.00_{2.68}$ & $80.87_{2.16}$ \\
\rowcolor{mygray} UNeXt (MICCAI'22)~\cite{unext}   & 1.47\,M   & 650.48 & 0.58  & $84.73_{1.23}$ & $91.20_{0.94}$ & $65.04_{2.71}$ & $74.16±2.84$
                                    & $71.04_{0.17}$ & $80.46_{0.16}$ \\
\rowcolor{mygray} EGE-Unet (MICCAI'23)~\cite{egeunet}  & 0.072\,M  & $303.08$ & $0.045$ & $84.72_{1.28}$ &$ 91.72_{0.75}$ & $58.90_{2.97}$ & $74.11_{2.34}$
                                    & $74.47_{0.43}$ & $85.36_{0.28}$\\
\rowcolor{mygray} ERDUnet (TCSVT'23)~\cite{erdunet} & 10.21\,M & 137.60 & 10.29 & $85.92_{1.50}$ & $92.01_{1.02}$  & $68.67_{1.90}$ & $77.74_{2.00}$ & $76.02_{0.40}$ & $84.63_{0.30}$  \\

\rowcolor{mygray} TinyU-Net (MICCAI'24)~\cite{tinyunet} & 0.48\,M & 359.50 & 1.66 & $85.67_{0.99}$ & $91.86_{0.80}$ & $66.21_{2.70}$ & $75.01_{2.60}$ & $74.03_{0.20}$ & $82.95_{0.10}$  \\

\hline
\rowcolor{myellow} Mobile U-ViT (Ours)  & 1.39\,M   & 326.24 & 2.51  & $87.28_{0.83}$ & \cellcolor{ll}{$92.90_{0.63}$} & \cellcolor{l}{$72.88_{2.72}$} & \cellcolor{l}{$81.18_{3.05}$}
                                       & \cellcolor{ll}{$77.70_{0.50}$} & \cellcolor{ll}{$85.90_{0.41}$}\\
\rowcolor{myellow} Mobile U-ViT-L (Ours)  & 7.88\,M  & 279.42  & 3.70 & \cellcolor{g}{\textbf{87.63}$_{0.91}$} & \cellcolor{g}{\textbf{93.13}$_{0.61}$} & \cellcolor{g}{\textbf{73.91}$_{2.65}$} & \cellcolor{g}{\textbf{82.16}$_{2.64}$}
                                       & \cellcolor{l}{$78.24_{0.38}$} & \cellcolor{l}{$86.37_{0.28}$} \\
\hline
\end{tabular}
}

\end{table*}

\renewcommand{\multirowsetup}{\centering}  
\begin{table}[h!]

\caption{Results on Dermoscopy and Polypscopy Datasets ($\text{Mean}_{\text{Std}}$).  Best results are highlighted as \colorbox{g}{\bf \!first\!}, \colorbox{l}{\!second\!} and \colorbox{ll}{\!third\!}. \colorbox{myellow}{Gray} indicates hybrid architecture method.  \label{tab:skinandpolpy}}
\resizebox{0.94\linewidth}{!}
{
\begin{tabular}{l| cc | cc}
\hline 

\multirow{3}{*}{Network} &  \multicolumn{2}{c|}{Dermoscopy} &  \multicolumn{2}{c}{Polypscopy}  \\
\cline{2-5}
& \multicolumn{2}{c|}{ISIC  (\%)} & \multicolumn{2}{c}{Kvasir  (\%)}  \\
\cline{2-5}
& IOU & F1 & IOU & F1 \\
\hline
\multicolumn{3}{l}{\textit{{\color{Gray} heavy-weight medical network}}} && \\

\rowcolor{mygray} U-Net (MICCAI'15)~\cite{unet}            & $82.18_{0.87}$ & $89.97_{0.52}$ & $86.52$ & $91.90$ \\
\rowcolor{mygray} CMU-Net (ISBI'23)~\cite{cmunet}          & $82.16_{1.06}$ & $89.92_{0.62}$ & \cellcolor{l}{89.02} & \cellcolor{l}{93.65}\\
\rowcolor{mygray} nnUNet (NM'21)~\cite{nnunet}             & \cellcolor{l}{$83.31_{0.59}$} & $89.84_{0.50}$ & $85.60$ & $91.50$\\
Swin-Unet (ECCV'22)~\cite{swinunet}                        & $82.15_{1.44}$ & $89.98_{0.87}$ & $73.58$ & $82.17$\\ 
\rowcolor{myellow} TransUnet (ArXiv'21)~\cite{transunet}   & $83.17_{1.25}$ & \cellcolor{g}{\textbf{90.57}$_{0.72}$} & $85.07$ & $90.62$\\  
\rowcolor{myellow} UCTransNet (AAAI'22)~\cite{uctransnet}  & $82.67_{0.40}$ & $89.39_{0.24}$ &\cellcolor{ll}{$87.97$} & \cellcolor{ll}{$92.93$}\\
 MissFormer (TMI'22)~\cite{missformer}  & $80.99_{0.28}$ & $88.03_{0.27}$ & $86.37$ & $91.79$ \\
                                    
\hline                                    
\multicolumn{3}{l}{\textit{{\color{Gray} light-weight natural network}}} && \\

\rowcolor{myellow} MobileViT-s (ICLR'21)~\cite{mobilevit}  & $80.12_{0.42}$ & $87.89_{0.43}$ & $84.21$ & $90.72$\\
\rowcolor{myellow} EdgeViT-s (ECCV'22)~\cite{edgevits} & $79.06_{0.56}$ & $87.06_{0.55}$ & 81.58 & 88.57\\

\rowcolor{myellow} EMO-6m (ICCV'23)~\cite{edgevits}   & $81.31_{0.73}$ & $88.74_{0.57}$ & 83.91 & 90.40 \\ 
\rowcolor{myellow} RepViT-m3 (CVPR'24)~\cite{repvit}  & $78.34_{0.61}$ & $86.62_{0.48}$ & $81.55$ & $88.75$\\
UniRepLKNet (CVPR'24)~\cite{unireplknet}              & $81.64_{0.24}$ &$88.82_{0.15}$ & $85.30$ & $91.18$ \\

\multicolumn{3}{l}{\textit{{\color{Gray} light-weight medical network}}} && \\

\rowcolor{myellow} MedT (MICCAI'21)~\cite{medt}       & $81.79_{0.94}$ & $89.74_{0.53}$ & $79.70$ & $87.40$\\
\rowcolor{mygray} UNeXt (MICCAI'22)~\cite{unext}      & $82.10_{0.88}$ & $89.93_{0.46}$ & $84.78$ & $91.01$\\
\rowcolor{mygray} EGE-Unet (MICCAI'23)~\cite{egeunet} & $82.19_{1.31}$ & \cellcolor{l}{$90.22_{0.79}$} & 77.41 & 85.70\\
\rowcolor{mygray} ERDUnet (TCSVT'23)~\cite{erdunet}   & $82.03_{0.20}$ & $88.96_{0.20}$ & $84.93$ & $90.88$ \\
\rowcolor{mygray} TinyU-Net (MICCAI'24)~\cite{tinyunet}  & $81.95_{0.30}$ & $88.99_{0.20}$ & $85.47$ & $91.35$ \\

\hline
\rowcolor{myellow} Mobile U-ViT (Ours)   & \cellcolor{ll}{$83.23_{0.39}$} & $89.86_{0.28}$ & 87.28 & 92.41\\
\rowcolor{myellow} Mobile U-ViT-L (Ours)  & \cellcolor{g}{\textbf{83.31}$_{0.36}$} & \cellcolor{ll}{$89.88_{0.25}$} & \cellcolor{g}{\textbf{89.07}} & \cellcolor{g}{\textbf{93.67}}\\
\hline
\end{tabular}
}

\end{table}

\renewcommand{\multirowsetup}{\centering}  
\begin{table}[h!]

\caption{Results on KiTs'23 Datasets.  Best results are highlighted as \colorbox{g}{\bf \!first\!}, \colorbox{l}{\!second\!} and \colorbox{ll}{\!third\!}.  \label{Tab.kits}}
\resizebox{1\linewidth}{!}
{
\begin{tabular}{l | c c c | c c c}
\hline 

\multirow{2}{*}{Network} & \multirow{2}{*}{Dice} & \multirow{2}{*}{HD95} & \multirow{2}{*}{ASD} & \multicolumn{3}{c}{KiTs'23 (Dice \%)}  \\
\cline{5-7}
 &  & & & Kidney & Tumor & Cyst \\
\hline

UNETR~\cite{unetr} & 56.91 & 28.11 & 12.96 & 91.40 & 54.60 & 24.74 \\
Swin UNETR~\cite{swinunetr} & 67.05 & 22.72 & 10.55 & \cellcolor{l}{93.31} & 71.53 & 36.30 \\
MedNeXt~\cite{mednext} & \cellcolor{ll}{70.30} & \cellcolor{l}{13.51} & \cellcolor{g}{\textbf{4.27}} & 92.93 & \cellcolor{ll}{74.33} & \cellcolor{ll}{43.63} \\
SegMamba~\cite{segmamba} & \cellcolor{g}{\textbf{71.52}} & 17.42 & 6.51 & \cellcolor{g}{\textbf{93.51}} & \cellcolor{l}{74.58} & \cellcolor{g}{\textbf{46.48}} \\
\hline
Mobile U-ViT (3D) & 68.49 & \cellcolor{ll}{15.33} & \cellcolor{ll}{5.98} & 92.60 & 73.14 & 39.73 \\
Mobile U-ViT-L (3D) & \cellcolor{l}{70.93} & \cellcolor{g}{\textbf{13.39}} & \cellcolor{l}{4.75} & \cellcolor{ll}{93.16} & \cellcolor{g}{\textbf{74.80}} & \cellcolor{l}{44.84} \\
\hline
\end{tabular}
}

\end{table}

\renewcommand{\multirowsetup}{\centering}  
\begin{table*}[h!]
\centering
\caption{Results on BTCV with five-fold in UNETR pipeline. Best results are highlighted as \colorbox{g}{\bf \!first\!}, \colorbox{l}{\!second\!} and \colorbox{ll}{\!third\!}. Note that GFLOPS and FPS are calculated for sub-volume crops (96 $\times$ 96 $\times$ 96). \label{Tab.btcv}}
\resizebox{1\linewidth}{!}
{
\begin{tabular}{l | c c c | c c c | ccccccccccc}
\hline
\multirow{2}{*}{Network} & \multirow{2}{*}{Params} & \multirow{2}{*}{GFLOPs} & \multirow{2}{*}{FPS} & \multirow{2}{*}{Dice} & \multirow{2}{*}{HD95} & \multirow{2}{*}{ASD} & \multicolumn{11}{c}{BTCV (Dice \%)} \\
\cline{8-18}
& & & & & & & \multirow{1}{*}{Spl} & \multirow{1}{*}{Kid} & \multirow{1}{*}{Gall} & \multirow{1}{*}{Eso} & \multirow{1}{*}{Liv} & \multirow{1}{*}{Sto} & \multirow{1}{*}{Aor} & \multirow{1}{*}{IVC} & \multirow{1}{*}{Veins} & \multirow{1}{*}{Pan} & \multirow{1}{*}{AG} \\
\cline{1-2}
\hline
UNETR (WACV'22)~\cite{unetr} & 92.61 & 82.70  & 52.00 & 71.67 & 10.92 & 3.32 & 86.14 & 82.73 & 55.99 & 67.23 & 92.50 & 73.80 & 85.26 & 77.59 & 59.81 & 53.71 & 53.75  \\ 
Swin UNETR (CVPR'22)~\cite{swinunetr} & 61.98 & 329.84 & 12.16 & 74.95 & 11.11 & 3.59 & 88.51 & 81.85 & 58.39 & \cellcolor{ll}{71.72} & \cellcolor{l}{94.18} & 78.47 & 86.97 & 80.51 & \cellcolor{l}{66.96} & 64.71 & \cellcolor{ll}{60.13}  \\ 
MedNeXt (MICCAI'23)~\cite{mednext} & 10.51 & 74.58 & 19.19 & \cellcolor{ll}{76.14} & \cellcolor{l}{8.57} & \cellcolor{ll}{2.78} & \cellcolor{l}{89.27} & \cellcolor{l}{83.88} & 60.85 & 71.66 & \cellcolor{g}{\textbf{94.30}} & 79.39 & \cellcolor{l}{88.94} & \cellcolor{ll}{82.25} & 64.85 & \cellcolor{ll}{70.63} & 59.94 \\
SegMamba (MICCAI'24)~\cite{segmamba} & 65.18 & 659.04 & 9.63 & \cellcolor{l}{76.54} & 10.47 & 3.52 & 88.19 & 82.11 & \cellcolor{ll}{61.52} & 70.54 & 93.29 & \cellcolor{ll}{80.13} & 87.24 & \cellcolor{g}{\textbf{82.81}} & \cellcolor{ll}{66.31} & \cellcolor{l}{72.38} & \cellcolor{g}{\textbf{64.23}}\\

\hline
Mobile U-ViT (3D) & 2.75 & 28.56 & 55.13 & 76.01 & \cellcolor{ll}{8.67} & \cellcolor{l}{2.63} & \cellcolor{ll}{88.89} & \cellcolor{ll}{83.53}  & \cellcolor{l}{64.02} & \cellcolor{l}{72.11} & 93.59 & \cellcolor{l}{80.45} & \cellcolor{ll}{88.75} & 82.07 & 64.94 & 67.09 & 59.58 \\

Mobile U-ViT-L (3D) & 11.06 & 110.88 & 28.24 & \cellcolor{g}{\textbf{78.42}} & \cellcolor{g}{\textbf{6.56}} & \cellcolor{g}{\textbf{2.18}} & \cellcolor{g}{\textbf{89.59}} & \cellcolor{g}{\textbf{85.81}} & \cellcolor{g}{\textbf{69.35}} & \cellcolor{g}{\textbf{72.20}} & \cellcolor{ll}{94.16} & \cellcolor{g}{\textbf{83.34}} & \cellcolor{g}{\textbf{89.40}} & \cellcolor{l}{82.60} & \cellcolor{g}{\textbf{68.85}} & \cellcolor{g}{\textbf{73.31}} & \cellcolor{l}{62.49} \\

\hline
\end{tabular}
}
\end{table*}

\begin{figure*}[htb]
  \centering
   \includegraphics[width=1\linewidth]{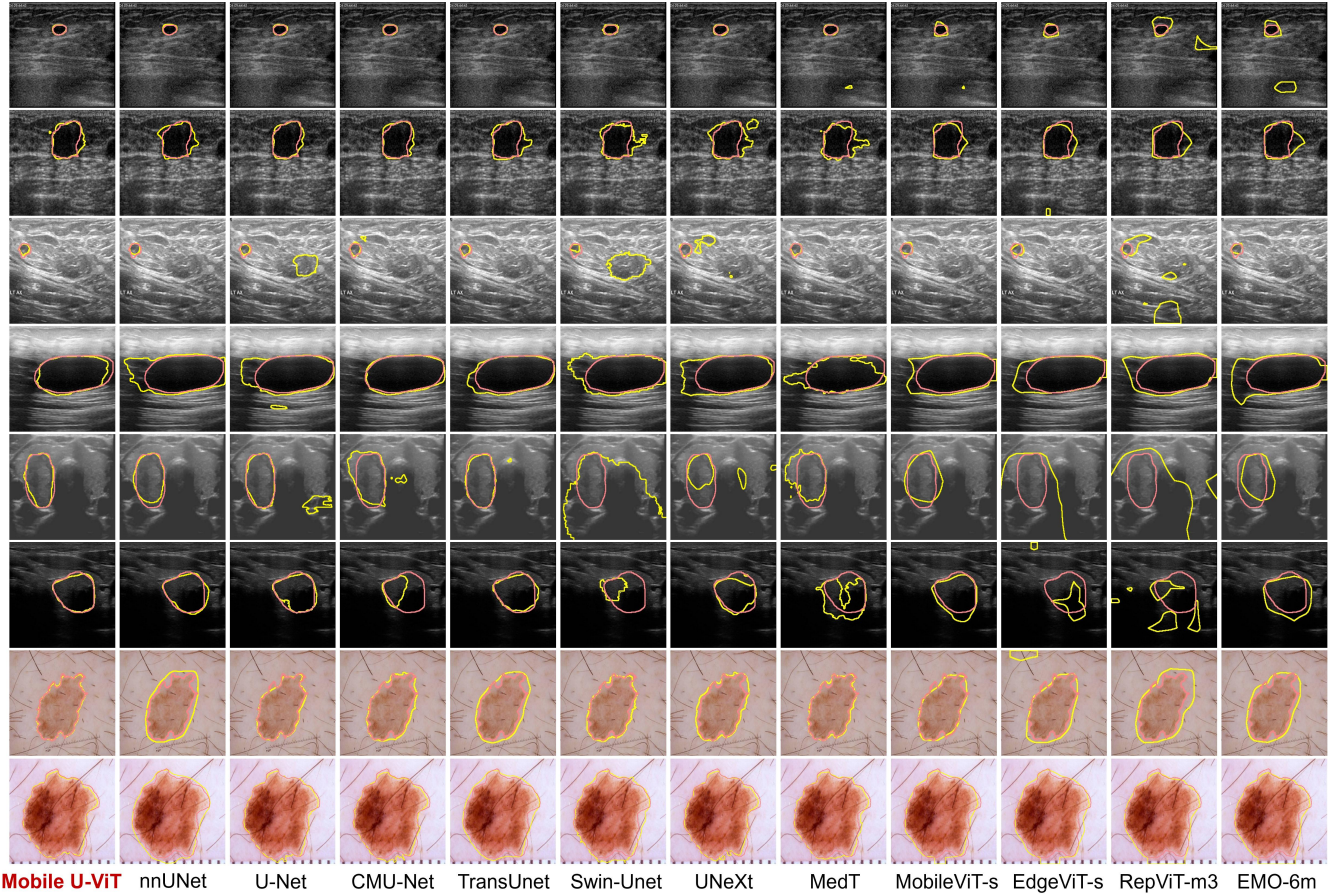}
   \caption{Visualization Results on Ultrasound and Dermoscopy Dataset.}
      \label{fig:appendix1}
\end{figure*}

\begin{figure*}[htb]
  \centering
   \includegraphics[width=1\linewidth]{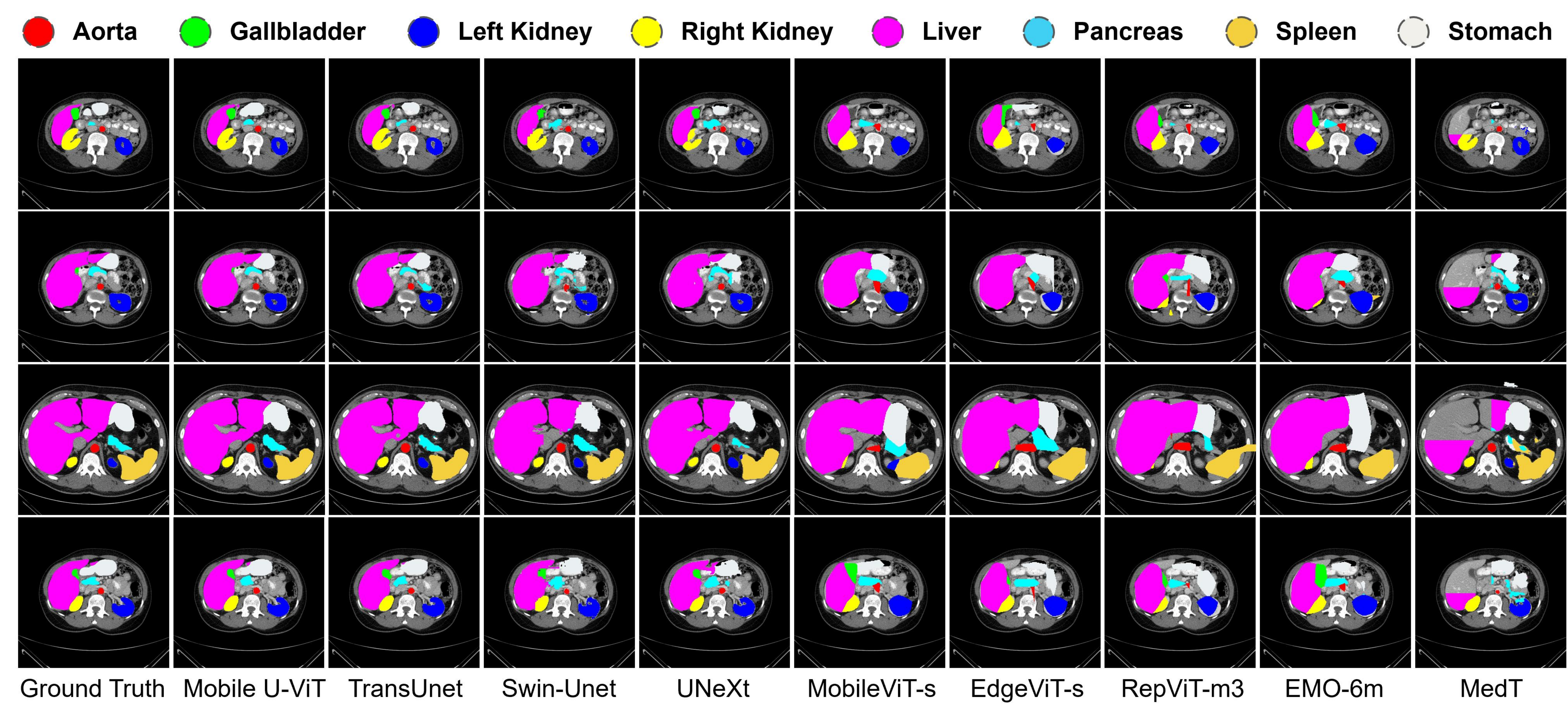}
   \caption{Visualization Results on Synapse Dataset.}
      \label{fig:appendix2}
\end{figure*}

\begin{figure}[t!]
  \centering
   \includegraphics[width=1\linewidth]{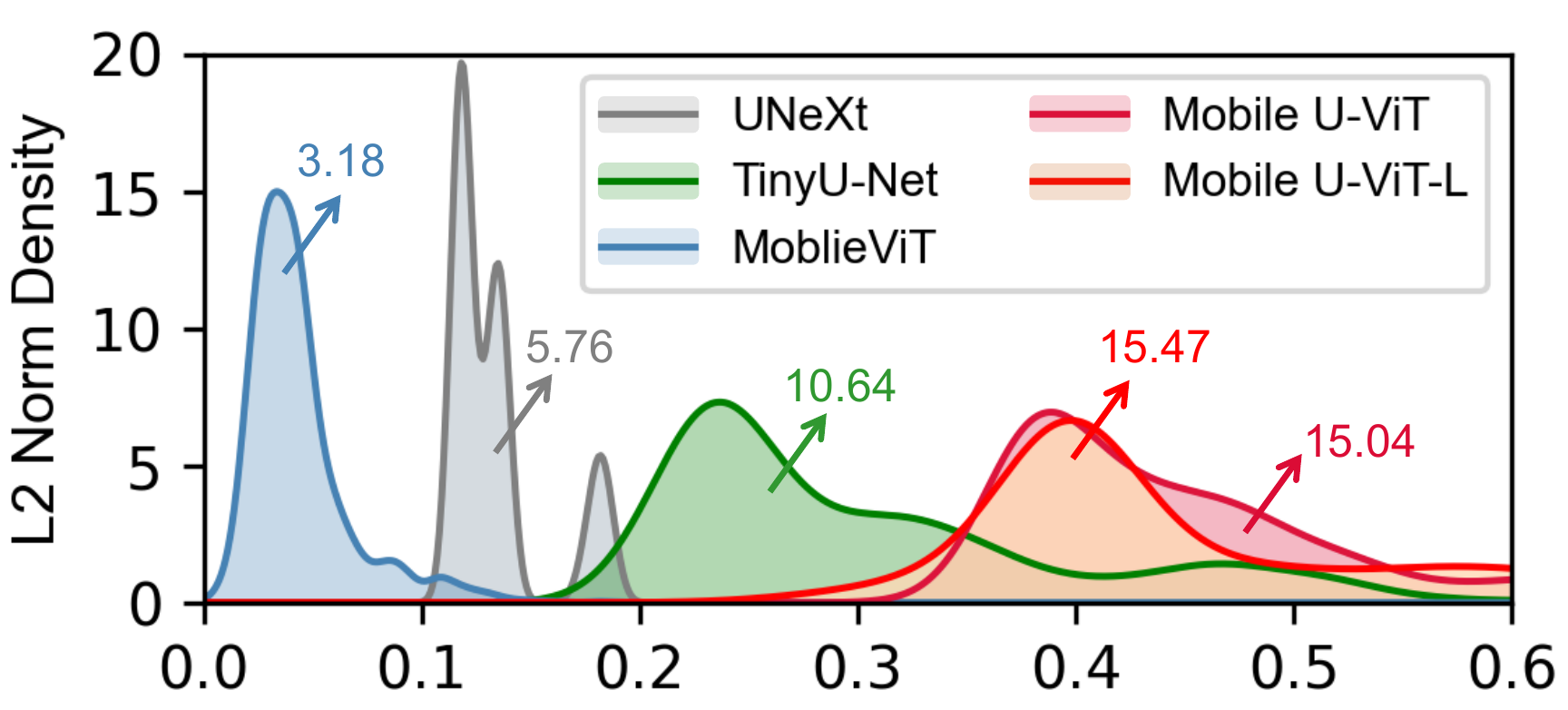}
   \vspace{-3mm}
   \caption{Channel redundancy is analyzed by the L2-norm of each channel, where high values represent low redundancy.}
      \label{fig:norm}
\end{figure}

\begin{figure*}[htb]
  \centering
   \includegraphics[width=1\linewidth]{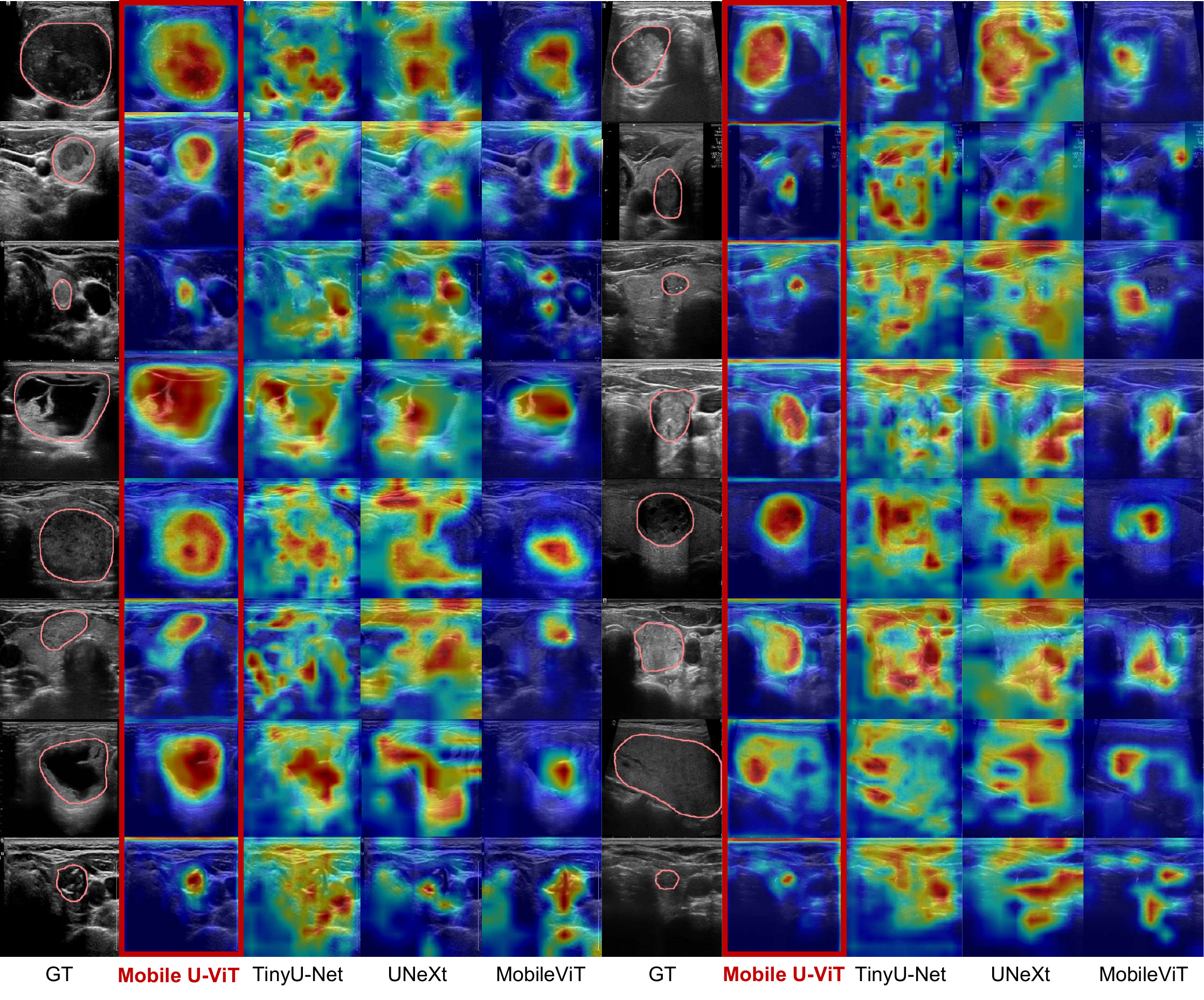}
   \caption{Visualization of Grad-CAM activations corresponding to the encoder outputs on BUS datasets. Our Mobile U-ViT effectively captures both global contextual information and fine-grained local details.}
      \label{fig:bus}
\end{figure*}

\begin{figure*}[htb]
  \centering
   \includegraphics[width=1\linewidth]{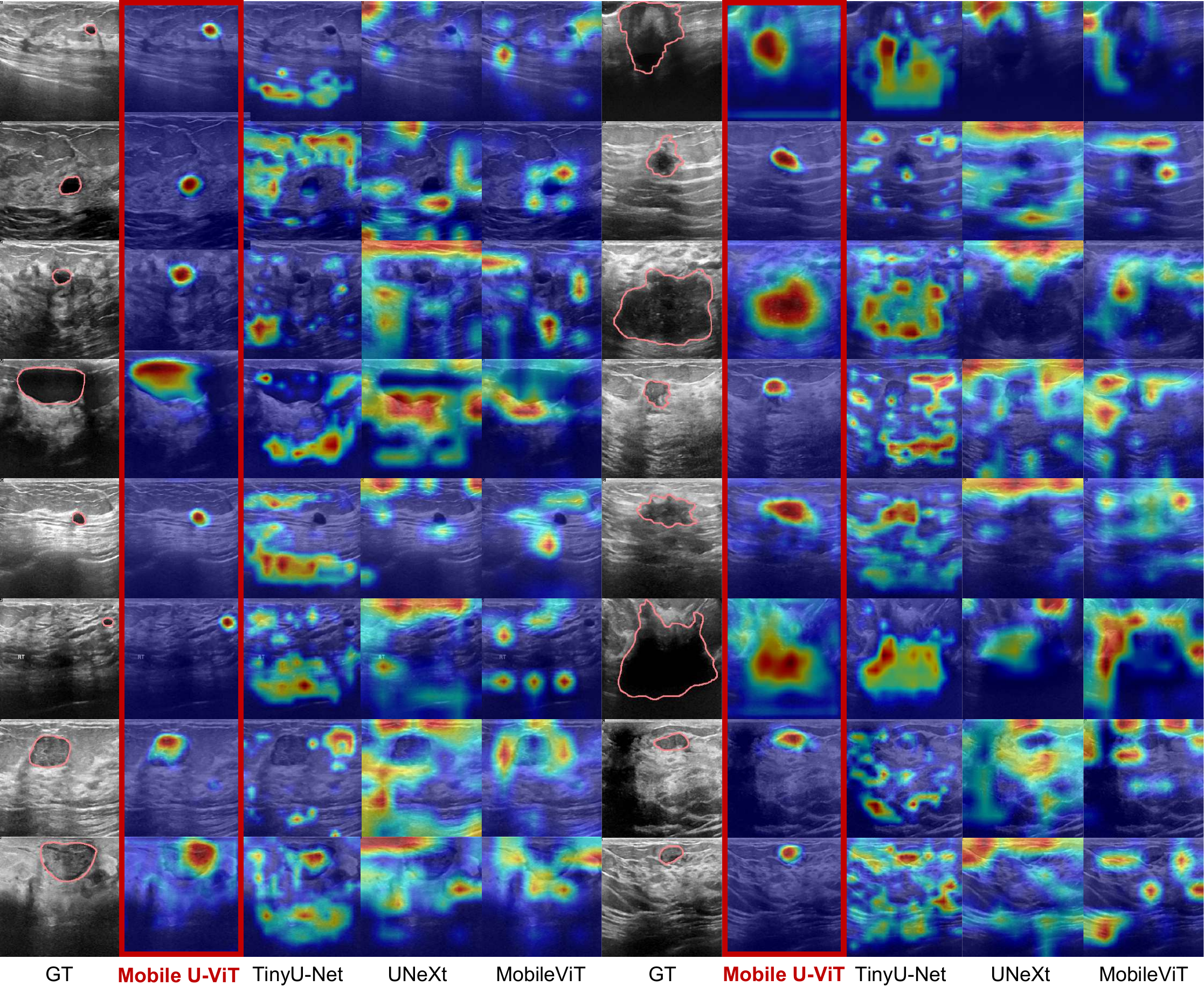}
   \caption{Visualization of Grad-CAM activations corresponding to the encoder outputs on BUSI datasets. Our Mobile U-ViT effectively captures both global contextual information and fine-grained local details.}
      \label{fig:busi}
\end{figure*}

\begin{figure*}[htb]
  \centering
   \includegraphics[width=1\linewidth]{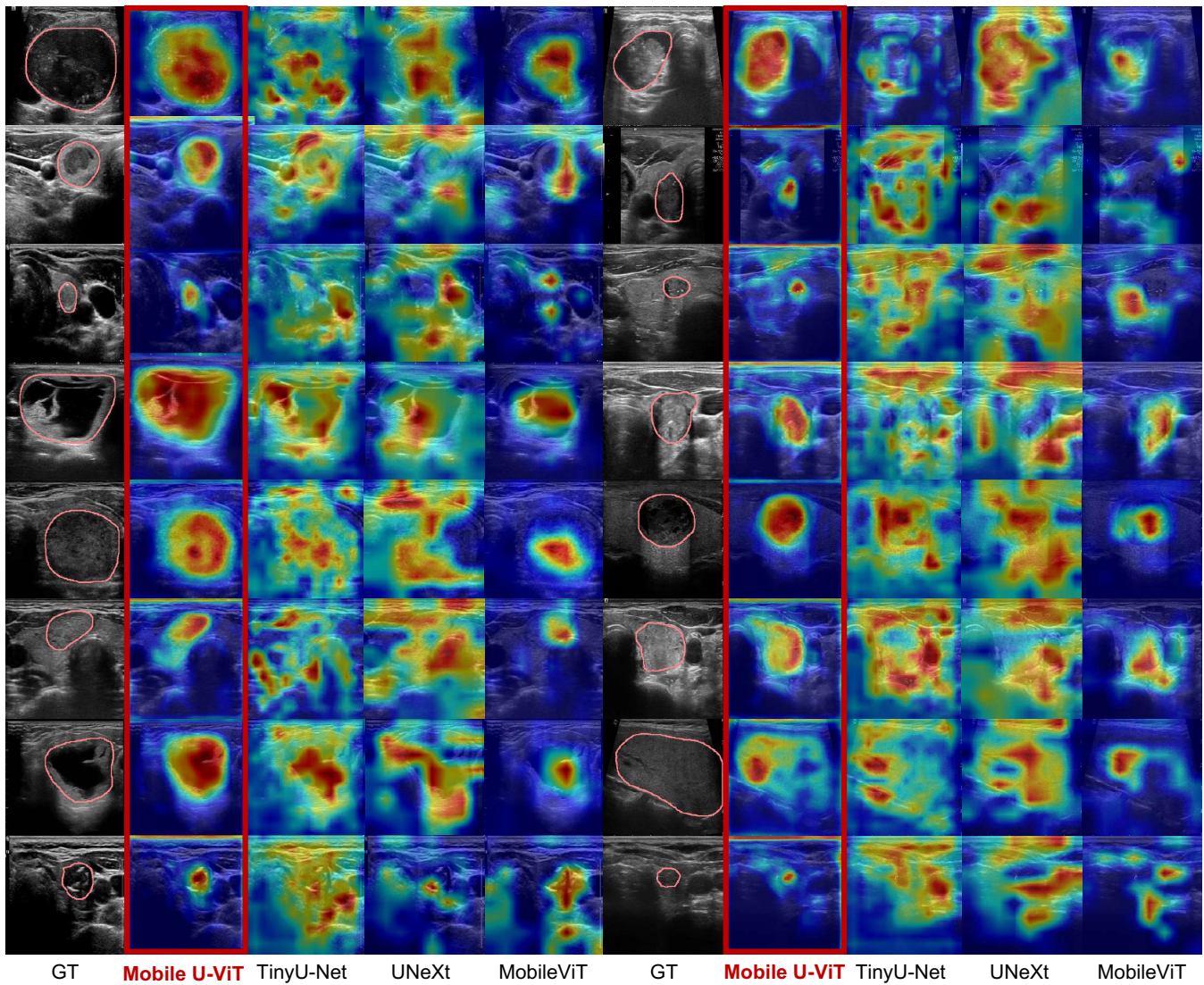}
   \caption{Visualization of Grad-CAM activations corresponding to the encoder outputs on TNSCUI datasets. Our Mobile U-ViT effectively captures both global contextual information and fine-grained local details.}
      \label{fig:tnscui}
\end{figure*}

\begin{figure*}[htb]
  \centering
   \includegraphics[width=1\linewidth]{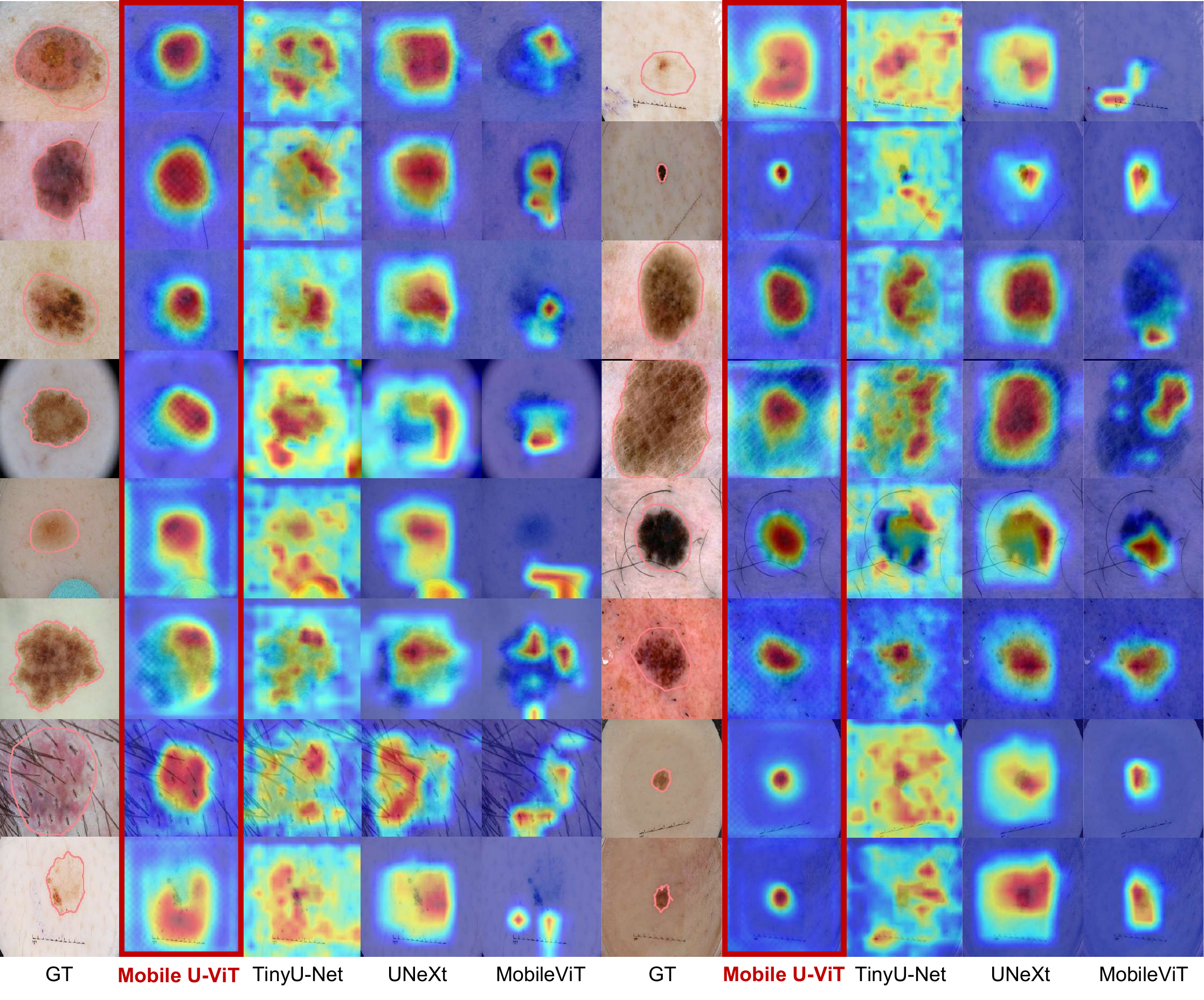}
   \caption{Visualization of Grad-CAM activations corresponding to the encoder outputs on ISIC datasets. Our Mobile U-ViT effectively captures both global contextual information and fine-grained local details.}
      \label{fig:isic}
\end{figure*}

\section{Implement details}

\subsection{Network Implement details}
For the upsampling process, we utilize bilinear interpolation, which helps preserve the finer details during the upsampling operation. The convolution layer within each stage employs a kernel size of $3\times3$, a stride of 1, and a padding of 1. For down-sampled skip-connection process, we employ two convolution operations (with a kernel size of 3, stride of 1, and padding of 1), followed by ReLU activation and batch normalization after each convolution.

\subsection{Experimental details}
We use a 70/30 split on the four 2D medical datasets (BUS, BUSI, TNSCUI, ISIC2018) for training and validation thrice. For kvasir, We use the official data partitioning. In addition, refer to previous work~\cite{transunet, swinunet}, we randomly split the Synapse dataset into 18 cases for training and 12 cases for validation. For BTCV, A five-fold cross-validation strategy is used to train models and select the best model in each fold. For KiTS'23, we use an 80/20 split for training and validation.

As shown in Tab.~\ref{tab:settings}, for 2D datasets, following~\cite{cmunet, unext}, we resize all training cases of five datasets to \( 256 \times 256 \) and apply random rotation and flip for simple data augmentations. In addition, we use the SGD optimizer with a weight decay of 1e-4 and a momentum of 0.9 to train all the networks. The initial learning rate is set to 0.01, and the poly strategy is used to adjust the learning rate. The batch size is set to
8 and the training epochs are 300. 

For 3D datasets, following~\cite{unetr, swinunetr}, the training datasets are interpolated to the isotropic voxel spacing of \( 1.5 \times 1.5 \times 2.0\) mm. Intensities are scaled to $[-175, 250]$, then normalized to $[0, 1]$. We sample the sub-volumes of $96 \times 96 \times 96$ voxels by ratios of positive and negative as 1:1 in 4 sub-crops. Augmentation probabilities for random flip, rotation, intensities scaling, and shifting are set to 0.5, 0.3, 0.1, 0.1, respectively. We train with the AdamW optimizer, an initial learning rate of 1e-4, and a cosine-annealing scheduler for all experiments. The experiments use a batch size of 4 ($1 \times 4$ sub-crops) and train the model for 100k steps.

\section{Experimental Results}
\subsection{Experiments on ultrasound images}
In ultrasound image segmentation tasks (BUS, BUSI, TNSCUI datasets). Our proposed Mobile U-ViT is compared to SOTA methods mentioned in Table~\ref{tab:ExOnUltrasound}. The experimental results demonstrate that our Mobile U-ViT achieves the best performance, striking a better balance between accuracy and computational cost. 

Specifically, in the BUS and BUSI datasets, nnUNet~\cite{nnunet} has competitive performances. However, our network achieves comparable results while maintaining a significantly smaller model size (1.39 M vs 26.10 M), and improved computational efficiency (2.51 GFLOPs vs 12.67 GFLOPs). Although nnUNet's performance is high, once we expand the network dimension, Mobile U-ViT-L achieves the best performance with an IoU score of 87.63 and 73.91 (0.1\% and 1.8\% higher than nnUNet). Even in the largest TNSCUI dataset, Mobile U-ViT obtains competitive performance while maintaining the minimum parameters.

Moreover, the light-weight CNNs like UNeXt~\cite{unext} and EGE-Unet~\cite{egeunet} do not yield satisfactory results. While their parameter and computation requirements significantly decreased their corresponding performance also declined. This phenomenon is also observed in natural networks. When we employ networks like MobileViT~\cite{mobilevit}, EdgeViT~\cite{edgevits}, RepViT~\cite{repvit}, EMO~\cite{emo}, EfficientViT\cite{efficientvit}, and UniRepLKNet\cite{unireplknet}, their effectiveness is limited as these networks are originally designed for specific tasks in natural images. When applied to medical images, their performance is much lower than Mobile U-ViT. On the other hand, networks that combine CNN and ViT, such as TransUnet~\cite{transunet} (105.32M parameters, 112.95 FPS, 38.52 GFLOPs) and UCTransNet~\cite{uctransnet} (66.24M parameters, 57.11 FPS, 32.98 GFLOPs), achieve certain levels of success but inevitably face a trade-off between performance and computational burden.

However, with the special design of encoder, our Mobile U-ViT maintains a nearly smallest light-weight model while achieving almost the best performance.  This indicates the effectiveness and correctness of our encoder's patch embedding and combining strategy CNNs with Transformers. Furthermore, Mobile U-ViT stands as the first successful model in making Transformer light-weight, setting a new benchmark in the medical domain.

\subsection{Experiments on dermoscopy and polypscopy images}
In the dermoscopy and polyp segmentation experiments, we focus on the challenging task of segmenting skin and polyp lesions under natural lighting conditions. As shown in Table~\ref{tab:skinandpolpy}, both Mobile U-ViT and Mobile U-ViT-L achieve the highest IoU scores for skin and polyp lesion segmentation. Notably, lightweight models such as EGE-Unet and TinyU-Net significantly reduce the parameter count and computational cost, while maintaining competitive accuracy. However, these improvements are largely attributed to the characteristics of dermoscopy and polypscopy images, which often contain rich textures, high contrast, and clearly defined edges—making the segmentation task relatively easier. Moreover, the high similarity between training and test images further simplifies the learning process, allowing most networks to fit the data effectively.

It is worth highlighting that nnUNet and TransUNet achieve top performance not only on ultrasound image datasets (BUS, BUSI, TNSCUI), but also on dermoscopy and polyp segmentation datasets (ISIC 2018 and Kvasir). This can be attributed to their strength as general-purpose medical segmentation networks—an ability our proposed Mobile U-ViT shares. Indeed, many medical segmentation tasks, such as those involving ultrasound or CT images, fall under the category of weak target segmentation. As shown in Table~\ref{tab:ExOnUltrasound}, most lightweight models suffer from substantial performance degradation, suggesting that their lightweight designs do not generalize well across diverse tasks. A truly effective lightweight model should maintain high performance across a wide range of medical segmentation scenarios—this is precisely where the value of our proposed network lies.

\subsection{Experiments on 3D CT volume} 

Our 3D extension of Mobile U-ViT demonstrates significant advantages across two challenging medical image segmentation benchmarks: BTCV and KiTS'23 (Table~\ref{Tab.btcv} and Table~\ref{Tab.kits}). On the BTCV dataset, Mobile U-ViT-L (3D) achieves superior segmentation accuracy compared to SegMamba, improving the Dice score by 1.88\% (78.42\% \textit{v.s.} 76.54\%) while requiring only 17\% of the parameters (11.06M vs. 65.18M).  The scalability of our approach is further evidenced on the larger KiTS'23 dataset, where Mobile U-ViT delivers robust tumor segmentation performance (74.80\% Dice) with precise boundary delineation (13.39 mm average HD95). These consistent results across different dataset scales and anatomical targets confirm that Mobile U-ViT effectively bridges the gap between computational efficiency and segmentation performance in 3D medical image analysis. We further test our model on \textcolor{red}{\textbf{MICCAI FLARE 2022 Playground AbdomenCT-1K: Fully Supervised Learning Benchmark}}, and Mobile U-ViT gains \underline{Top 6} positions on the leader-board.

\subsection{Analysis}

\subsection{Statistic Analysis of Encoded Feature} To investigate the representational capacity of our method, we compute the L2 norm of encoder output feature maps across different models. As shown in Fig.~\ref{fig:norm}, Mobile U-ViT consistently yields higher L2 norm values compared to prior baselines, indicating stronger activation and richer semantic responses. This suggests that our model is capable of capturing more discriminative features, which also reflects a better utilization of network parameters, aligning with feature foreground performance that more informative foreground features are associated with higher energy in the representation space.

To further interpret the behavior of our model, we visualize the Grad-CAM activations derived from the encoder outputs. As shown in Fig~\ref{fig:bus},~\ref{fig:busi},~\ref{fig:tnscui} and~\ref{fig:isic}, Mobile U-ViT produces more focused and semantically aligned activation maps compared to other light-weight models. In particular, the Grad-CAM heatmaps of our model demonstrate a higher degree of overlap with the target foreground regions, suggesting that the encoder is more effectively attending to task-relevant features.

Notably, even in challenging scenarios with complex backgrounds or ambiguous boundaries (e.g. Thyroid ultrasound images in Fig~\ref{fig:tnscui}), Mobile U-ViT maintains precise attention to the object of interest. This aligns well with the segmentation masks generated by our model, further confirming the strong correspondence between the internal feature representations and the desired output.

\subsection{Qualitative Visualization}

Figure~\ref{fig:appendix1} and~\ref{fig:appendix2} show the visualization results across multiple medical imaging modalities. For 2D segmentation, Mobile U-ViT effectively captures finer details and better aligns with the ground truth, showing fewer mismatches and more precise delineation of structures. In slice-base multi-organ segmentation tasks, Mobile U-ViT-L demonstrates superior segmentation performance, successfully identifying structures and boundaries with minimal error. These results underline the robustness and accuracy of Mobile U-ViT in handling diverse and complex medical datasets, showcasing its capability to perform well in multiple-modalities segmentation tasks while maintaining high efficiency.

\bibliographystyle{ACM-Reference-Format}
\bibliography{sample-base}

\end{document}